\documentclass[
reprint,
superscriptaddress,
amsmath,amssymb,
aps,
prb,
]{revtex4-2}

\usepackage[T1]{fontenc}
\usepackage[latin9]{inputenc}
\usepackage{mathrsfs}
\usepackage{amsmath}
\usepackage{amssymb}
\usepackage{graphicx}
\usepackage{xcolor}
\usepackage{braket}



\usepackage{hyperref} 

\begin{document}
\preprint{}

\title{Generalized atomic limit of a double quantum dot coupled to superconducting leads}

\author{Martin \v{Z}onda}
\affiliation{Department of Condensed Matter Physics, Faculty of Mathematics and Physics, Charles University, Ke Karlovu 5, CZ-12116 Praha 2, Czech Republic}

\author{Peter Zalom}
\affiliation{Institute of Physics, Czech Academy of Sciences, Na Slovance 2, CZ-18221 Praha 8, Czech Republic}

\author{Tom\'a\v{s} Novotn\'{y}}
\affiliation{Department of Condensed Matter Physics, Faculty of Mathematics and Physics, Charles University, Ke Karlovu 5, CZ-12116 Praha 2, Czech Republic}

\author{Georgios Loukeris}
\affiliation{Institute of Physics, Albert Ludwig University of Freiburg, Hermann-Herder-Strasse 3, DE-79104 Freiburg, Germany}

\author{Jakob Bätge}
\affiliation{Institute of Physics, Albert Ludwig University of Freiburg, Hermann-Herder-Strasse 3, DE-79104 Freiburg, Germany}

\author{Vladislav Pokorn\'{y}}
\affiliation{Institute of Physics, Czech Academy of Sciences, Na Slovance 2, CZ-18221 Praha 8, Czech Republic}

\begin{abstract}

We present an exactly solvable effective model of a double quantum dot coupled to superconducting leads. This model is a generalization of the well-known superconducting atomic limit approximation of the paradigmatic superconducting impurity Anderson model. However, in contrast to the standard atomic limit and other effective models, it gives quantitatively correct predictions for the quantum phase transition boundaries, subgap bound states as well as Josephson supercurrent in a broad range of parameters including experimentally relevant regimes. The model allows fast and reliable parameter scans important for the preparation and analysis of experiments which are otherwise inaccessible by more precise but computational heavy methods such as quantum Monte Carlo or the numerical renormalization group. The scans also allowed us to identify and investigate new previously unnoticed phase diagram regimes. We provide a thorough analysis of the strengths and limitations of the effective model and benchmark its predictions against numerical renormalization group results.
      
\end{abstract}
\maketitle
%
%%%%%%%%%%%%%%%%%%%%%%%%%%%%%%%%%%%%%%%%%%%%%%%%%%%%%%%%%%%
\section{Introduction}
\label{intro}

The recent progress in controlled fabrication of systems that combine nanodevices containing few active orbitals with superconducting reservoirs brought a multitude of tunable heterostructures~\cite{DeFranceschi2010,Benito2020}. The examples range through systems of magnetic adatoms on superconducting surfaces~\cite{Yazdani1997,Heinrich2018}, weak links \cite{Snyder2018}, small scale single and multiple quantum dots (QDs)~\cite{DeFranceschi2010,Delagrange2015,Delagrange2016,Saldana2018} to island structures~\cite{Saldana2022} attached to superconducting leads. In general, the overall complexity of these experimental systems steadily increases. The surface experiments are already probing atomic dimers~\cite{Ruby2018,Choi2018, Kamlapure2018, Ding2021, Beck2021, Kuster2021}, weak links have been prepared in multi-terminal-lead arrangements~\cite{Draelos2019,Pankratova2020} and tunable double quantum dots (DQDs) have been constructed in serial~\cite{Saldana2018,Saldana2020,Rasmussen2018} and in parallel configurations~\cite{Vekris2021,Steffensen2022}. This progress is motivated by the ability of these heterostructures to probe basic physical concepts as well as by the proposed applications in future electronics, computational devices and sensors~\cite{DeFranceschi2010,Benito2020}.

A common feature and crucial characteristic of the superconducting heterostructures is the existence of bound states within the gap of the superconductor. Although they are of the same physical origin, depending on the parameter regime and the physical realization of the investigated system, they are referred to either as Yu-Shiba-Rusinov (YSR) or Andreev bound states (ABS)~\cite{Zitko2022}, as preferred here. Moreover, in both cases crossing of these bound states at the Fermi energy marks a  quantum phase transition (QPT). A singlet-doublet transition of this kind is known as the $0-\pi$ transition~\cite{Oguri2004,Grove2007,Jorgensen2007,Tanaka2007-jpsj,Maurand2012,Pillet2013,Delagrange2015,Delagrange2016,Delagrange2018} in experiments with single-dot Josephson junctions. There the underlying QPT manifests itself by a sudden change of the sign of the measured supercurrent.
   
Unfortunately, reliable theoretical investigations of the evolution of ABS in realistic multi-parametric space often require prohibitively expensive numerical approaches such as the numerical renormalization group (NRG)~\cite{Bulla2008,Yoshioka2000,Bauer2007,Yao2014,Zitko2015dd,Zitko2016,Zalom2021a,Zalom2021b} or various types of quantum Monte Carlo (QMC) ~\cite{Siano2004,Luitz2010,Delagrange2015,Pokorny2021}.
This limitation can be partially sidestepped by controlled analytic approximations, e.g., various mean-field approaches~\cite{Yeyati1997,Rodero1999,Yoshioka2000,Rodero2011}, perturbation expansions~\cite{Alastalo1998,Vecino2003,Meng2009,Zonda2015,Zonda2016} or functional renormalization group techniques~\cite{Karrasch2008,Wentzell2016}. However, their validity range is often not sufficient for typical experiments or they cannot capture all of the relevant regimes. Another frequently employed strategy is to utilize simple effective models, such as the zero-bandwidth approximation (ZBW)~\cite{Vecino2003,Rasmussen2018} or the superconducting atomic limit (AL) effective model~\cite{Meden2019}. Their great advantage is that they can be straightforwardly extended to complex scenarios while remaining exactly solvable. However, they are limited either to qualitative descriptions or require an \emph{ad hoc} reparametrization to match the experiments or full numerical solutions~\cite{Meden2019,Zitko2015dd}. Moreover, even then they frequently predict transport properties which differ by orders of magnitude from the exact results or perturbative calculations~\cite{Rasmussen2018}.

In our paper we present a remedy for these shortcomings. We introduce an effective model which is based on an effective AL Hamiltonian with scaled parameters. For brevity we call it GAL (or MGAL) model, because it is devised to reproduce the approximation of the phase boundary position known as the generalized atomic limit (GAL)~\cite{Zonda2015,Zonda2016} or its modification (MGAL) according to  Ref.~\cite{Kadlecova2019}.
GAL was derived perturbatively in the on-dot Coulomb interaction but it can be also justified by other procedures~\cite{Pokorny2022}. Originally, it was limited to the case of single QD and provided only the position of the phase boundaries. Similar approximations for more complex setups have been missing. In contrast, the GAL model introduced here has the form of an exactly solvable AL finite-dimensional Hamiltonian with all its advantages, e.g., it can be used to calculate ABS energies or Josephson current. On top of this, unlike AL, it is in a good quantitative agreement with the NRG data for many relevant regimes.

A big advantage of our effective GAL model is its scalability to more complex systems. Interestingly, GAL model then mitigate most of the shortcomings observed for single QD. We show and utilize this feature for the case of a serial double quantum dot (SDQD) coupled to two superconducting leads. We demonstrate that GAL (unlike AL or ZBW) correctly predicts the position of the singlet-doublet phase boundaries, the subgap energy spectrum (ABS) and the Josephson current not only qualitatively but also quantitatively in a broad range of parameters including experimentally relevant regimes. This makes GAL a useful tool not only for fast preliminary scans of a broad parameter space but also for direct analysis of experiments.  Moreover, it can be straightforwardly generalized to longer chains of QDs and even more complex structures while keeping the advantage of a relatively small Hilbert space even in comparison with other effective models such as ZBW.
Thus it can be utilized in theoretical investigations of complicated setups which present a serious challenge for both NRG and QMC, and where the standard AL gives only qualitative results~\cite{Domanski2017,Zienkiewicz2019,Gorski2018}.

The paper is structured as follows: In Sec.~\ref{sec:Ham} we introduce the Anderson impurity models for single QD and SDQD systems, which is then followed in Sec.~\ref{sec:AL} by their corresponding AL effective models. In Sec.~\ref{sec:SIS} we briefly summarize the main results of the GAL approximation for single QD. At the beginning of Sec.~\ref{sec:DIS} we introduce the GAL model for SDQD and its modification MGAL for the away from half-filling case. We then discuss the GAL predictions for the phase diagrams (Sec.~\ref{sec:PD}), subgap states (\ref{sec:ABS}) and Josephson current (\ref{sec:JC}) at half-filling with comparison to NRG results. In Sec.~\ref{sec:Prel}, we present a detailed MGAL scan of phase boundaries for the case away from half-filling. In Sec. ~\ref{sec:AHF} we test a region predicted by MGAL, in which a small change of model parameters leads to dramatic evolution of phase diagrams, via NRG and also benchmark MGAL for experimentally relevant parameters. Section~\ref{sec:summary} gives a summary of the main results. Some of the technical details related to the Green functions, GAL, MGAL and NRG as well as additional supporting analysis of phase diagrams and subgap spectra are postponed to the corresponding Appendices.

%%%%%%%%%%%%%%%%%%%%%%%%%%%%%%%%%%%%%%%%%%%%%%%%%%%%%%%%%%%
\section{Model}
\label{sec:Ham}

\begin{figure}[tbh]
\includegraphics[width=1.0\columnwidth]{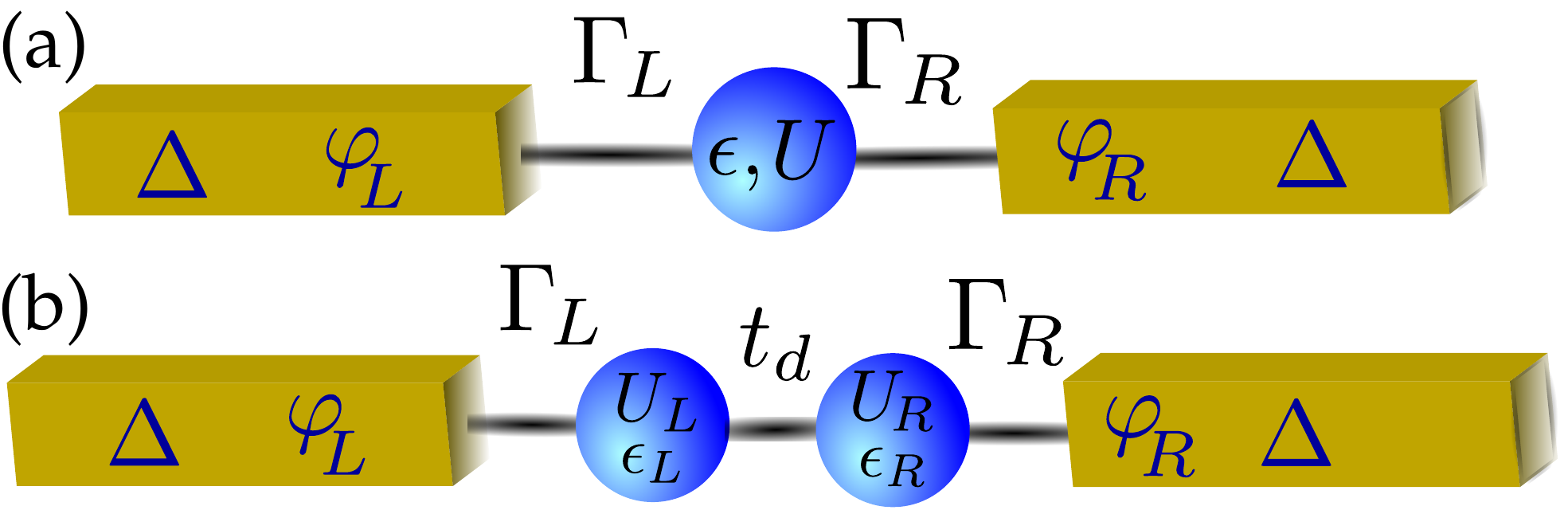}
\caption{
	Illustration of the model of a single quantum dot (a) and a serial double quantum dot (SDQD) (b) connected to two superconducting leads.
}
\label{fig:setup}
\end{figure}

The paradigmatic model for quantum dots coupled to superconducting leads is the superconducting impurity Anderson model (SCIAM)~\cite{Rodero2011,Meden2019}. Its general Hamiltonian can be written as
\begin{equation}
	\mathcal{H}=\mathcal{H}^\text{imp}+\sum_{j=L,R}\left(\mathcal{H}^\text{lead}_{j}+\mathcal{H}^\text{hyb}_{j}\right),
	\label{eq:Ham}
\end{equation}
where $\mathcal{H}^\text{int}$ describes one or more impurities in a serial configuration as sketched in Fig.~\ref{fig:setup}.
In the case of single dot it reads 
\begin{eqnarray}
\mathcal{H}^\text{imp}_{\text{1d}}
&=&
\epsilon\sum_{\sigma}
d_{\sigma}^\dag d_{\sigma}^{\phantom{\dag}}
+U d_{\uparrow}^\dag d_{\uparrow}^{\phantom{\dag}}
 d_{\downarrow}^\dag d_{\downarrow}^{\phantom{\dag}}
\\ 
&=&\varepsilon \sum_{\sigma}
\left(d_{\sigma}^\dag d_{\sigma}^{\phantom{\dag}}-\frac{1}{2}\right)\nonumber\\
&&+\frac{U}{2} \left(d_{\uparrow}^\dag d_{\uparrow}^{\phantom{\dag}}+
 d_{\downarrow}^\dag d_{\downarrow}^{\phantom{\dag}}-1\right)^2 + \text{const}.,
\label{eq:Dot}
\end{eqnarray}
where $d_{\sigma}^\dag$ ($d_{\sigma}$) creates (annihilates) an electron with spin $\sigma$ on the impurity with energy $\epsilon$ and $U$ is the local Coulomb interaction (charging energy) on the dot. 
In the case of SDQD the impurity part becomes
\begin{eqnarray}
\mathcal{H}^\text{imp}_{\text{2d}}&=&\sum_{j\sigma}\epsilon_{j} 
d_{j\sigma}^\dag d_{i\sigma}^{\phantom{\dag}}
-t_d\sum_\sigma \left(d_{L\sigma}^\dag d_{R\sigma}^{\phantom{\dag}}
+\textrm{H.c.}\right)\label{eq:Dotsa}\\
&&+\sum_j U_j d_{j\uparrow}^\dag d_{j\uparrow}^{\phantom{\dag}} 
d_{j\downarrow}^\dag d_{j\downarrow}^{\phantom{\dag}}\nonumber\\
&=&\sum_{j\sigma}\varepsilon_{j} 
\left(d_{j\sigma}^\dag d_{i\sigma}^{\phantom{\dag}}-\frac{1}{2}\right)
-t_d\sum_\sigma \left(d_{L\sigma}^\dag d_{R\sigma}^{\phantom{\dag}}+\textrm{H.c.}\right)\nonumber\\
&&+\sum_j \frac{U_j}{2} \left(d_{j\uparrow}^\dag d_{j\uparrow}^{\phantom{\dag}}+ 
d_{j\downarrow}^\dag d_{j\downarrow}^{\phantom{\dag}}-1\right)^2 + \text{const}..
\label{eq:Dots}
\end{eqnarray}
Here, $d_{j\sigma}^\dag$ creates an electron on the site $j=L,R$ with spin $\sigma$ and energy $\epsilon_j$, $t_d$ is the inter-dot hopping amplitude and $U_j$ is the local Coulomb interaction on the site $j$. Note that we have shifted both Hamiltonians by a constant term and introduced shifted energy levels $\varepsilon_{j} \equiv \epsilon_{j}+U/2$ measured with respect to the particle-hole symmetric point (half-filling)~\cite{Zitko2015dd}. 

The second term in Hamiltonian~\eqref{eq:Ham} describes left and right superconducting leads according to the BCS theory via
\begin{equation}
\label{eq:Leads}
\begin{aligned}
\mathcal{H}_j^\text{lead}&=
\sum_{\mathbf{k}\sigma}\varepsilon_{j\mathbf{k}}
c_{j\mathbf{k}\sigma}^\dag c_{j\mathbf{k}\sigma}^{\phantom{\dag}}\\
&-\Delta_j\sum_\mathbf{k}\left(e^{i\varphi_j}
c_{j\mathbf{k}\uparrow}^\dag c_{j\mathbf{-k}\downarrow}^\dag+\textrm{H.c.}\right),
\end{aligned}
\end{equation}
where $c_{j\mathbf{k}\sigma}^\dag$ creates an electron with spin $\sigma$ 
and energy $\varepsilon_{j\mathbf{k}}$ in the lead $j \in L,R$ and 
$\Delta_{j}e^{i\varphi_j}$ is the complex superconducting order parameter.
In the following we assume $\Delta_L=\Delta_R\equiv\Delta$, which is the typical case in experimenal realizations, and introduce phase difference $\varphi=\varphi_L-\varphi_R$ where $\varphi_L=\varphi/2$ and $\varphi_R=-\varphi/2$ without loss of generality~\cite{Meden2019}.

The last term in Hamiltonian~\eqref{eq:Ham} describes the hybridization between the central part and leads:
\begin{equation}
\mathcal{H}_j^\text{hyb}=
t_{j}\sum_{\mathbf{k}\sigma}\left(c_{j\mathbf{k}\sigma}^\dag 
d_{j\sigma}^{\phantom{\dag}}+\textrm{H.c.}\right),
\label{eq:Hubs}
\end{equation}
where $t_j$ is the hopping between the lead $j=L,R$ and the
neighboring quantum dot. In our analysis we assume the tunnel-coupling magnitudes $\Gamma_j(E)=\pi |t_{j}|^2\sum_\mathbf{k}\delta(E-\varepsilon_{j\mathbf{k}})$
to be constant. Moreover, in the case of single QD, one can focus solely on the symmetric coupling ($\Gamma_L=\Gamma_R=\Gamma=\Gamma_T/2$), because all typical observables of the asymmetric scenario ($\Gamma_L\neq\Gamma_R$), including the Josephson current, can be easily extracted from the symmetric case~\cite{Kadlecova2017}.

In the paper we also apply a convention of omitting the subscript whenever an equivalent magnitude of parameters on the
left and right side of the heterostructure is present, i.e., $\Gamma\equiv\Gamma_L=\Gamma_R$, $U\equiv U_L= U_R$ and $\varepsilon \equiv\varepsilon_L=\varepsilon_R$. If not stated otherwise, we use $\Delta$ as the energy unit.

%%%%%%%%%%%%%%%%%%%%%%%%%%%%%%%%%%%%%%%%%%%%%%%%%%%%%%%%%%%
\subsection{Atomic limit}
\label{sec:AL}

Utilizing standard equation-of-motion technique~\cite{Novotny2005}, and taking the limit of infinite bandwidth followed by the limit of infinite superconducting gap $\Delta\rightarrow\infty$ allows to define an effective AL model of the SCIAM~\cite{Meden2019}. Although it does not reflect any experimentally relevant regime, it often gives a correct qualitative picture~\cite{Tanaka2007,Bauer2007,Meng2009,Eldridge2010,Trocha2015}. For single QD, the AL model reads
\begin{eqnarray}
	\mathcal{H}^{\text{AL}}_{\text{1d}}=&\varepsilon& \sum_{\sigma}
	\left(d_{\sigma}^\dag d_{\sigma}^{\phantom{\dag}}-\frac{1}{2}\right)
	+\frac{U}{2} \left(d_{\uparrow}^\dag d_{\uparrow}^{\phantom{\dag}}+
	d_{\downarrow}^\dag d_{\downarrow}^{\phantom{\dag}}-1\right)^2\nonumber\\
	&+&\left(\Delta_{\varphi} d_{\uparrow}^\dag d_{\downarrow}^\dag + \mathrm{H.c.}\right),
	\label{eq:DotAL}
\end{eqnarray}
where $\Delta_\varphi = \Gamma_L e^{-i \varphi/2}+\Gamma_R e^{i \varphi/2} = \Gamma_T \cos(\varphi/2)$ for $\Gamma_L=\Gamma_R$.
An analogous procedure for SDQD leads to AL Hamiltonian~\cite{Zitko2015dd}:
\begin{equation}
\label{eq:DotsAL}
\begin{aligned}
	\mathcal{H}^\text{AL}_{\text{2d}}&=\sum_{j\sigma}\varepsilon_{j}
	\left(d_{j\sigma}^\dag d_{i\sigma}^{\phantom{\dag}}-\frac{1}{2}\right)
	-t_d\sum_\sigma \left(d_{L\sigma}^\dag d_{R\sigma}^{\phantom{\dag}}
	+\textrm{H.c.}\right) \\
	&+\sum_j \frac{U_j}{2} \left(d_{j\uparrow}^\dag d_{j\uparrow}^{\phantom{\dag}}+
	d_{j\downarrow}^\dag d_{j\downarrow}^{\phantom{\dag}}-1\right)^2 \\
	&+ \sum_j \left(\Gamma_j e^{i\Phi_j}d_{j\uparrow}^\dag d_{j\downarrow}^\dag + \mathrm{H.c.}\right).\\
\end{aligned}
\end{equation}
While AL models are useful for qualitative analysis, they show several drawbacks. For example, in the case of single dot the position of QPT given by
\begin{equation}
\left(\frac{U}{2}\right)^2 = \varepsilon^2 + \Gamma^2_T\cos^2\frac{\varphi}{2}
\label{eq:AL_boundary}
\end{equation}
does not reproduce the NRG or QMC results. To match the precise numerical data, a significant shift of model parameters, often very far away from the original ones, is necessary. A more serious problem is related to the Josephson current. For a single dot, the AL model predicts an $U$-independent current in the singlet phase and zero current in the doublet phase~\cite{Meden2019}.
Neither of these predictions is supported by the full SCIAM solutions or experiments. This issue cannot be tamed by any manipulation of the model parameters as it is a consequence of the absence of the incoherent band states in AL model~\cite{Pokorny2022}. However, as we show in our paper, these drawbacks are largely eliminated in the GAL model for SDQDs.

%%%%%%%%%%%%%%%%%%%%%%%%%%%%%%%%%%%%%%%%%%%%%%%%%%%%%%%%%%%
\section{GAL model for single dot system}
\label{sec:SIS}
 
To find the GAL model, we start with the results of the perturbation theory in $U$ according to Refs.~\cite{Zonda2015,Zonda2016}. As shown in the cited works, the energies of the lowest ABSs follow $\omega_{ABS}\approx F(\hat{G}_0,U)/(1+\Gamma_T/\Delta)$ in the vicinity of the QPT. The functional $F(\hat{G}_0,U)$ depends on the non-interacting ($U=0$) Green function $\hat{G}_0$  of the full model and on the interaction $U$ and smoothly passes through zero exactly at the QPT point. After omitting band contributions this property can be used to obtain analytical formula for the approximate position of the phase boundaries,
\begin{equation}
\left(\frac{U}{2(1+\Gamma_T/\Delta)}\right)^2 = \varepsilon^2 + \Gamma_T^2\cos^2\dfrac{\varphi}{2},
\label{eq:GAL_SD}
\end{equation}
which at half-filling follows the NRG results closely up to surprisingly strong Coulomb interaction ($U\approx 10\Delta$)~\cite{Zonda2015,Zonda2016,Kadlecova2017,Kadlecova2019}. Interestingly, although derived by different means, formula~\eqref{eq:GAL_SD} clearly resembles the AL result~\eqref{eq:AL_boundary} with a correction for finite superconducting gap, therefore, it is called the \emph{generalized} AL.

\begin{figure}
	\includegraphics[width=1.0\columnwidth]{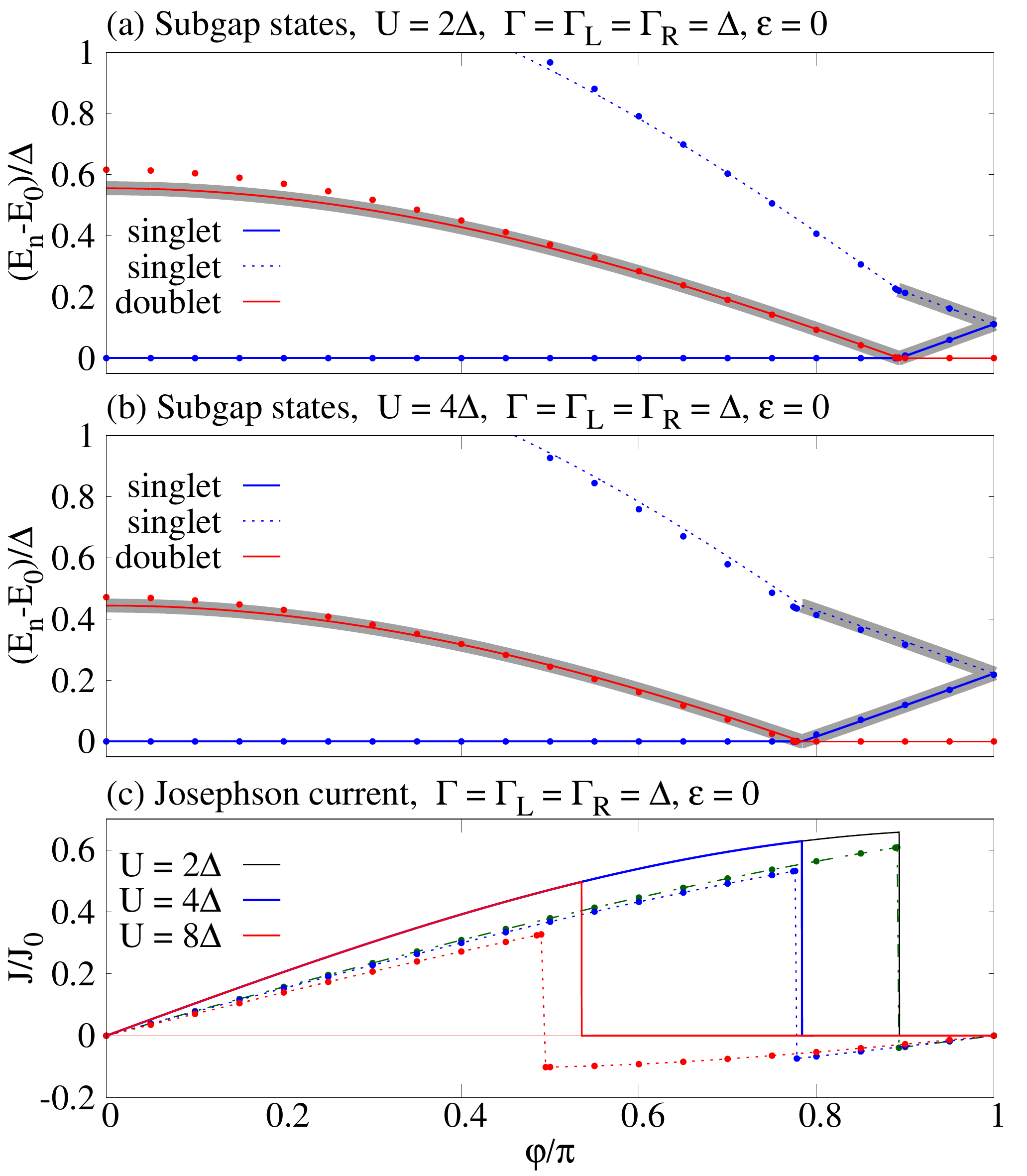}
	\caption{
		Comparison of NRG (circles) and GAL (solid and dashed lines) results for the single quantum dot at half-filling coupled to two superconducting leads. 
		[(a) and (b)] The difference between the energy of excited subgap many-body states and the ground state as a function of $\varphi$ for $\Gamma=\Delta$, $U=4\Delta$ (a) and $U=8\Delta$ (b). Note that because of the parity selection rules only the excited states underscored by the gray stripes will form ABS visible in the single-particle excitation spectra.   
		(c) Phase-dependent Josephson current for $\Gamma=\Delta$, $U=2$, $4$ and $8\Delta$, where $J_0=e\Delta/\hbar$. 
		\label{fig:SD_ABS}}
\end{figure} 

Nevertheless, the connection between GAL~\eqref{eq:GAL_SD} approximation and the actual AL model was not considered so far. Yet, there is a clear link. To show this, it is enough to take the AL Hamiltonian~\eqref{eq:DotAL} and subject it to three requirements. First, we require the AL QPT boundary to follow the GAL formula~\eqref{eq:GAL_SD}. Second, the ABS should follow $\approx 1/(1+\Gamma_T/\Delta)$ in the vicinity of QPT. Third, the shifted energy levels should be zero at the half-filling (particle-hole symmetric point). All this can be achieved by the following simple scaling of the AL model parameters
\begin{eqnarray}
\varepsilon 
&\rightarrow&
\tilde\varepsilon
=\nu\varepsilon,
\label{eq:GAL_trans_eta}
\\
\Delta_\varphi 
&\rightarrow&
\tilde\Delta_\varphi=\nu\Delta_\varphi,
\\
U &\rightarrow&
\tilde U=\nu^2 U
\label{eq:GAL_trans_U},
\end{eqnarray}
where $\nu=1/(1+\Gamma_T/\Delta)$ is a scaling factor reintroducing the finite superconducting gap into the AL model. Note, that the original energy level, therefore, scales as
\begin{equation}
\epsilon \rightarrow \tilde \epsilon=\nu\left[\epsilon + (1-\nu)\frac{U}{2}\right]
\label{eq:GAL_trans_eps}
\end{equation} 
and at half-filling we have $\tilde \epsilon = -\tilde U/2$ as expected.

In other words, a parameter of the AL Hamiltonian~\eqref{eq:DotAL} is scaled by $\nu$ if it multiplies a quadratic term and by $\nu^2$ if it belongs to a quartic term.
Strictly speaking, rigorous derivation of this scaling is still missing. Nevertheless, it can be justified by a mapping used originally for the derivation of the microscopic basis for the Fermi liquid theory~\cite{Nozieres-1998} as recently shown in Ref.~\cite{Pokorny2022}. 

The resulting rescaled effective AL Hamiltonian is what we refer to as the GAL model for brevity. It has the form of~\eqref{eq:DotAL} and is, thus, exactly solvable. Its spectrum consists of one doublet state and two singlets. The eigenenergies are zero for the doublet state while for the two singlets they read
\begin{equation} 
E=\frac{U}{2(1+\Gamma_{T}/\Delta)^2} \pm \frac{\sqrt{\varepsilon^2+\Delta^2_\varphi}}{1+\Gamma_{T}/\Delta}.
\label{eq:gse}
\end{equation}
As required, this reproduces the GAL formula~\eqref{eq:GAL_SD} for the position of the phase boundary. 

\begin{figure}
	\includegraphics[width=1.0\columnwidth]{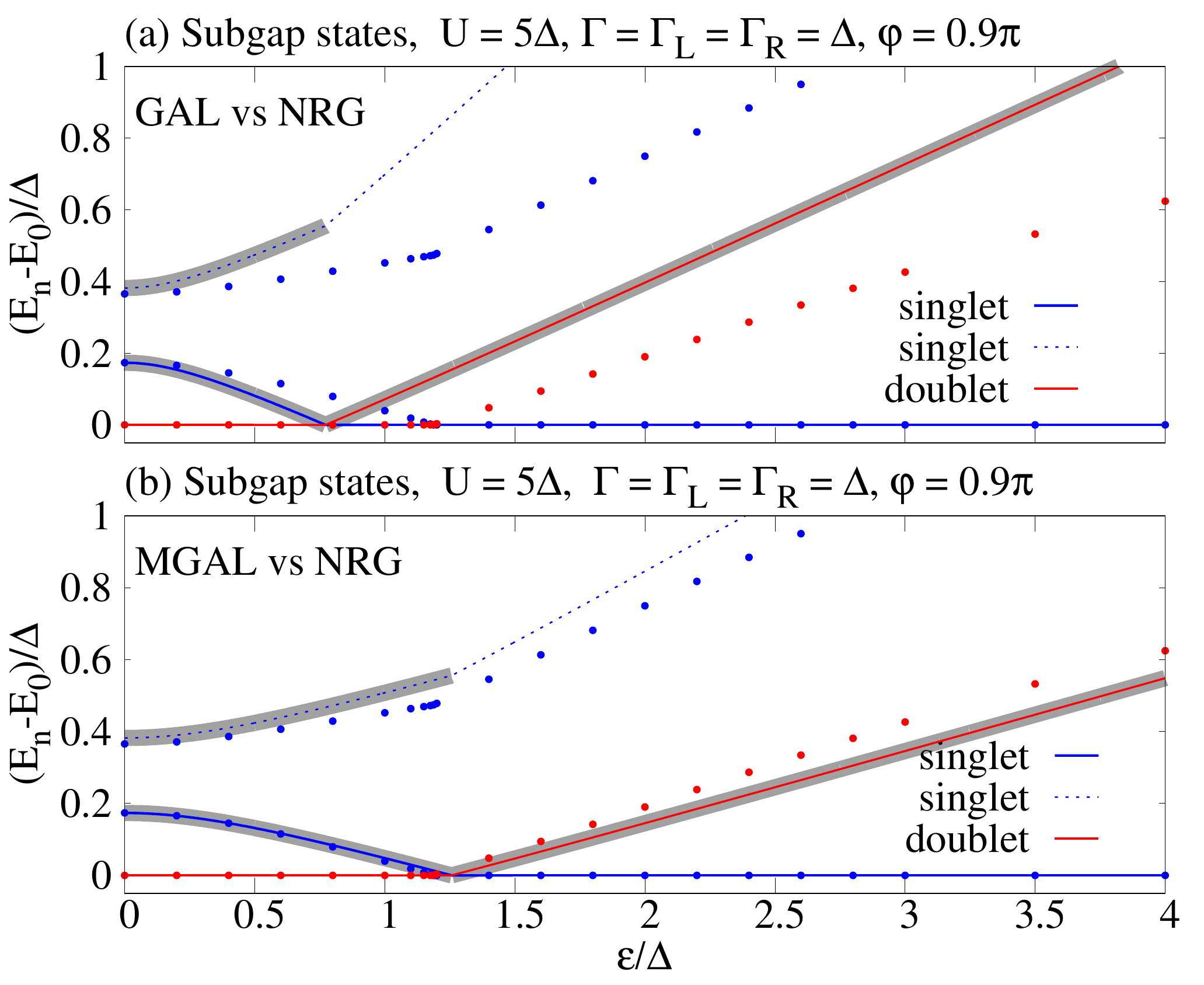}
	\caption{
		The difference between the energy of excited subgap many-body states and the ground state as a function of $\varepsilon$ for $\Gamma=\Delta$, $U=5\Delta$ and $\varphi=0.9\pi$. The two panels show the comparison of data calculated with NRG (circles) and GAL (solid and dashed lines) in (a) and MGAL in (b). Because of the parity selection rules only the excited states underscored by the gray stripes will form ABS visible in the single-particle excitation spectra.   
		\label{fig:SD_ABSb}}
\end{figure}

The GAL formula~\eqref{eq:GAL_SD} is in a good agreement with the position of the phase boundaries obtained via NRG calculations only near the half-filling condition ($\varepsilon=0$). A much better
agreement away from half-filling can be obtained by introducing a phenomenological scaling of the local energy level~\cite{Kadlecova2019} known as MGAL which replaces Eq.~\eqref{eq:GAL_trans_eta} by
\begin{equation} 
	\varepsilon \rightarrow \tilde\varepsilon_{\textrm{MGAL}}=\nu^2\varepsilon\sqrt{1+\frac{2\Gamma_T}{\nu U}}.
	\label{eq:epsMGAL}
\end{equation}
To distinguish this case we refer to an effective model where Eq.~\eqref{eq:epsMGAL} is used as the MGAL model. Because GAL and MGAL are identical at half-filling ($\varepsilon=0$) we utilize phenomenological MGAL only when $\varepsilon \neq 0$.  

Having the GAL Hamiltonian, the subgap energy spectrum can be easily obtained as shown in Figs.~\ref{fig:SD_ABS}(a),~\ref{fig:SD_ABS}(b) and \ref{fig:SD_ABSb}~(a). The GAL model rectifies the AL energy spectrum. For half-filled case this leads to a solid agreement with the NRG states even far away from the QPT. Away from half-filling GAL deviates from the NRG as illustrated in Fig.~\ref{fig:SD_ABSb}~(a). However, this deviation can be to a large extend corrected by using MGAL as shown in Fig.~\ref{fig:SD_ABSb}~(b).   

Yet, the GAL model does not solve all of the AL shortcomings. Since the Josephson current at zero temperature is given by $J=2e/\hbar(\partial E_0/\partial \varphi)$~\cite{Meden2019}, where $E_0$ is the ground-state energy, we obtain
\begin{equation}
J=J_0\frac{\Gamma_T^2 \sin\varphi}{2(\Delta+\Gamma_T)\sqrt{\varepsilon^2 + \Gamma_T^2 \cos^2(\varphi/2)}}
\end{equation}    
with $J_0 = e\Delta/\hbar$ for the singlet phase and zero for the doublet phase as its ground state energy does not depend on $\varphi$. The GAL model leads to quantitative improvement of the Josephson current in the singlet phase as shown by the comparison with the NRG data in Fig.~\ref{fig:SD_ABS}(c). Considering the perturbative origin of GAL, it is not surprising that the discrepancies increase with $U$, yet, we get a reasonable agreement unless $U\gg\Gamma$.
However, the GAL model inherits from the AL approximation both the already mentioned qualitative drawbacks. The Josephson current is zero in the doublet phase and in the singlet phase does not depend on $U$. These issues can be solved by introducing a simple band correction~\cite{Pokorny2022}. This correction incorporates some effects of the leads, neglected by the superconducting atomic limit, and restores the continuous part of the impurity spectral function above the gap, as discussed in detail in Ref.~\cite{Pokorny2022}. In the case of double QDs, the effects of such correction are much weaker than for single dots. More importantly, both above-mentioned issues of GAL related to Josephson current are naturally rectified in SDQDs without the necessity to include such correction at all. Therefore, we avoid it as we focus solely on SDQDs in the rest of the paper. 

%%%%%%%%%%%%%%%%%%%%%%%%%%%%%%%%%%%%%%%%%%%%%%%%%%%%%%%%%%%
\section{GAL and MGAL models for serial double dot}
\label{sec:DIS}

%-------------------------Fig3----------------------
\begin{figure}[tbh]
	\includegraphics[width=1.00\columnwidth]{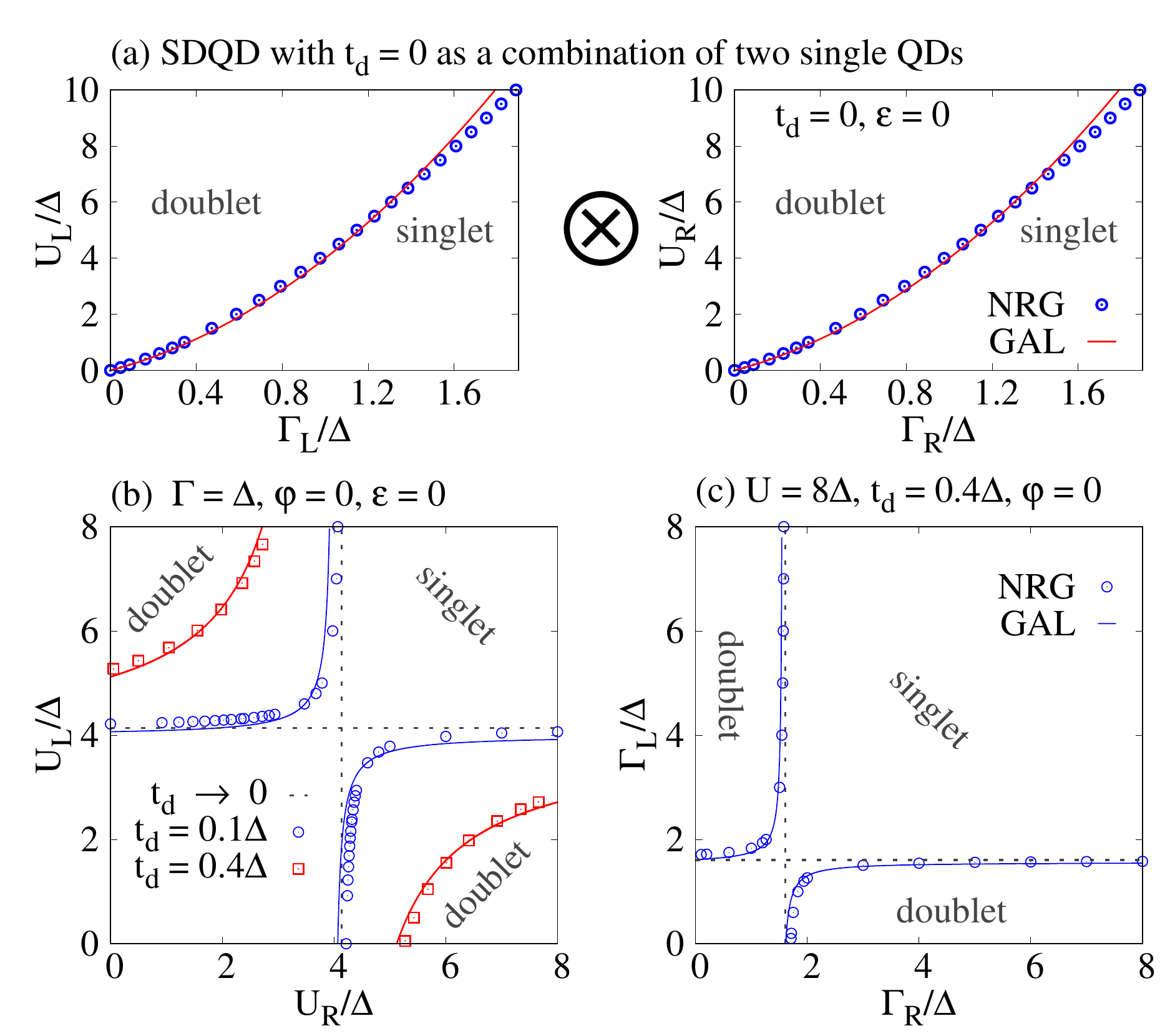}
	\caption{
		(a) The SDQD system at $t_d=0$ can be decoupled into two single dot subsystems. The resulting ground-state phase diagram is then a simple tensor-product-like combination of the corresponding single impurity phase diagrams with a solid agreement between NRG (points) and GAL (lines).
		(b)
		Ground-state phase diagrams of the SDQD system at $\Gamma=\Delta$ and $\varphi=0$. Two finite values of $t_d$ are presented at varying $U_L$ and $U_R$. The panel (a) explains the position of the $t_d \rightarrow 0$ phase boundaries plotted by a dashed line.
		(c)
		Ground-state phase diagrams of the SDQD system at parameters from the experimental study published in Ref.~\cite{Saldana2018}. The symbols represent NRG results and lines of corresponding colors are the GAL predictions in all panels.
		\label{fig:UU_PhD}}
\end{figure}

Adapting GAL scaling~\eqref{eq:GAL_trans_eta}-\eqref{eq:GAL_trans_eps} to a more complex system of SDQD requires some caution due to the emergence of new terms in AL Hamiltonian~\eqref{eq:DotsAL} and the fact that there is no GAL formula to guide us here. We thus start with the limiting case $t_d=0$ for which the system becomes a combination of two Hamiltonians describing independent single QD each connected to its own lead through $\Gamma_j$. In such a case, we naturally generalize the single dot GAL scaling to
\begin{eqnarray}
\varepsilon_j \rightarrow \tilde\varepsilon_j=\nu_j\varepsilon_j,
\label{eq:GALDQD_trans_eta}
\\
\Gamma_j \rightarrow \tilde\Gamma_j=\nu_j\Gamma_j,
\\
U_j \rightarrow \tilde{U}_j=\nu_j^2U_j,
\end{eqnarray}
where $\nu_j=1/(1+\Gamma_j/\Delta)$. When $t_d \neq 0$, we use
\begin{equation}
t_d \rightarrow \tilde t_d=\sqrt{\nu_L\nu_R}t_d
\label{eq:GALDQD_trans_td}
\end{equation}
since it multiplies a quadratic term in the SDQD AL Hamiltonian. Analogously to the single dot case, it proved to be advantageous to use MGAL for the away from the half-filling scenarios where instead of Eq.~\ref{eq:GALDQD_trans_eta} we have
\begin{equation} 
	\varepsilon_j \rightarrow \tilde\varepsilon_j^{\textrm{MGAL}}=\nu_j^2\varepsilon_j\sqrt{1+\frac{2\Gamma_j}{\nu_j U_j}}.
	\label{eq:DQDepsMGAL}
\end{equation}
Although at half-filling MGAL reduces to GAL we strictly distinguish these two models in the following discussion to stress the phenomenological nature of MGAL. Consequently, we use MGAL only away from half-filling.   

The GAL and MGAL Hamiltonians for the SDQD system have a form of Eq.~\eqref{eq:DotsAL} but with rescaled model parameters. Such Hamiltonian can be divided into its singlet, doublet and triplet subspaces~\cite{Zitko2015dd}. This allows a straightforward and trivial numerical diagonalization. Moreover, analytical solutions are possible for some limiting but useful cases (see Appendix~\ref{Sec:AppGALH}). Consequently, calculating GAL phase diagrams takes seconds on any modern PC while equivalent solutions of the full SCIAM model via NRG or QMC can be computationally very demanding. Yet, the GAL results are often in excellent agreement with these much more elaborated techniques. We show this in the following sections where we compare GAL with the NRG solutions of the SCIAM.
The GAL model can reliably predict phase boundaries, complicated energy dependencies of the subgap states (ABS) and even the Josephson current. Such a fast and simple tool has a lot of benefits. For example a broad parametric scan via MGAL model, which would not be feasible with NRG, allowed us to find previously unadressed regimes of SDQDs.

%%%%%%%%%%%%%%%%%%%%%%%%%%%%%%%%%%%%%%%%%%%%%%%%%%%%%%%%%%%
\subsection{Phase diagrams at half-filling}
\label{sec:PD}

It is illustrative to start the discussion with the $t_d=0$ case, which decouples into a tensor-product-like combination of two single dot subsystems. Each subsystem is identical to a single dot coupled to its respective lead. Increasing $t_d$ adiabatically for $\varphi=0$ allows us to combine the two, not necessarily equal, dots. The state of each of them is located in its corresponding single dot phase diagram shown in Fig.~\ref{fig:UU_PhD}(a). This leads to three possible ground states.

First, we can combine two dots from (single-dot) singlet ground-state phase regions each. The resulting combined ground state is, therefore, a singlet. The second possibility is to choose such $U_L$ and $U_R$ (or $\Gamma_L$ and $\Gamma_R$) that both belong to (single-dot) doublet state regions. For $t_d=0$ this leads to a degenerated singlet-triplet combination. However, for any finite $t_d$ the singlet-triplet is split by the inter-dot exchange coupling of $\approx 4 t_d^2/(U_L+U_R)$ (see Ref.~\cite{Zitko2015dd} and Appendix~\ref{Sec:AppGALH}) and the ground-state is therefore again a singlet. Consequently, if we combine equal dots, the combined ground-state is always a singlet for $\varphi=0$. For the sake of clarity we refer to the double-dot singlet that emerges (for $t_d\rightarrow 0$) due to the combination of two singlets as type I and to the one that combines two doublets as type II. Lastly there is a third option where one of the dots comes from singlet and the other from doublet (single-dot) region. This leads to a combined doublet ground-state. The resulting phase boundaries for $t_d\rightarrow0$ at varying $U_j$ or $\Gamma_j$ respectively are shown by dashed lines in Figs.~\ref{fig:UU_PhD}(b) and~\ref{fig:UU_PhD}(c). They reflect the tensor-product-like combinations of two single-dot phase diagram of Fig.~\ref{fig:UU_PhD}(a) with the three choices discussed above.

Naturally, increasing $t_d$ changes this simple picture and modifies the phase diagrams~\cite{Zitko2015dd}. Nevertheless, the GAL model can account for this change.
To be specific, the two singlet phases merge and push the doublet phases to higher asymmetries between $U_L$ and $U_R$ or $\Gamma_L$ and $\Gamma_R$ as seen both from the effective GAL model (solid lines) as well as from the NRG results for SCIAM (symbols) in Figs.~\ref{fig:UU_PhD}(b) and~\ref{fig:UU_PhD}(c). Let us point out that for the shown parameters the GAL model is in a very good agreement with the NRG for a fraction of computational costs.
%----------fig-4-----------
\begin{figure}[tbh]
\includegraphics[width=1.0\columnwidth]{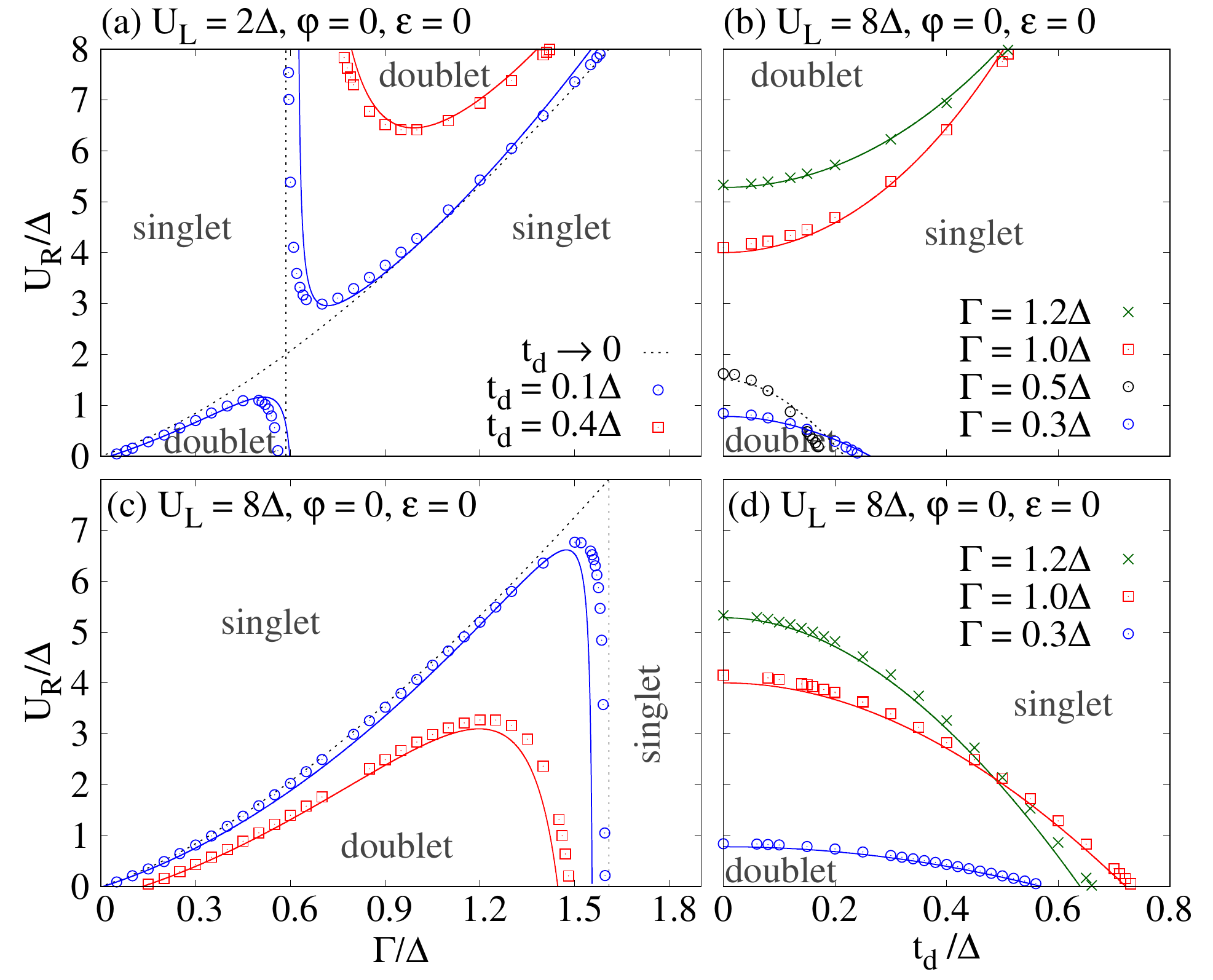}
\caption{
	(a)
	Phase diagrams of the half-filled case with $\Gamma_L = \Gamma_R = \Gamma$, $U_L=2\Delta$ and varying $U_{R}$ and $\Gamma$ at $\varphi=0$. Symbols represent NRG results, lines of corresponding colors are GAL predictions. Two finite values of $t_d$ are presented together with $t_d\rightarrow 0$ case (dashed lines).
	(b)
	$t_d$ dependence of the case in (a) for three selected values of $\Gamma=\Gamma_L=\Gamma_R$.
	(c)
	The same as in (a) only $U_L=8\Delta$.
	(d)
	The same as in (b) only $U_L=8\Delta$.
}
\label{fig:UG_PhD}
\end{figure}

The parameters in Fig.~\ref{fig:UU_PhD}(c) had been taken from the
experimental work presented in Ref.~\cite{Saldana2018}. Despite strong interaction $U=8\Delta$ the GAL predictions for SDQD are still correct. Its usefulness is underscored by the fact that the difference between NRG and GAL boundaries are below the resolution of a typical experiment~\cite{Maurand2012,Delagrange2015,Delagrange2016}.
  
%------------fig-5--------
\begin{figure}[tbh]
\includegraphics[width=1.0\columnwidth]{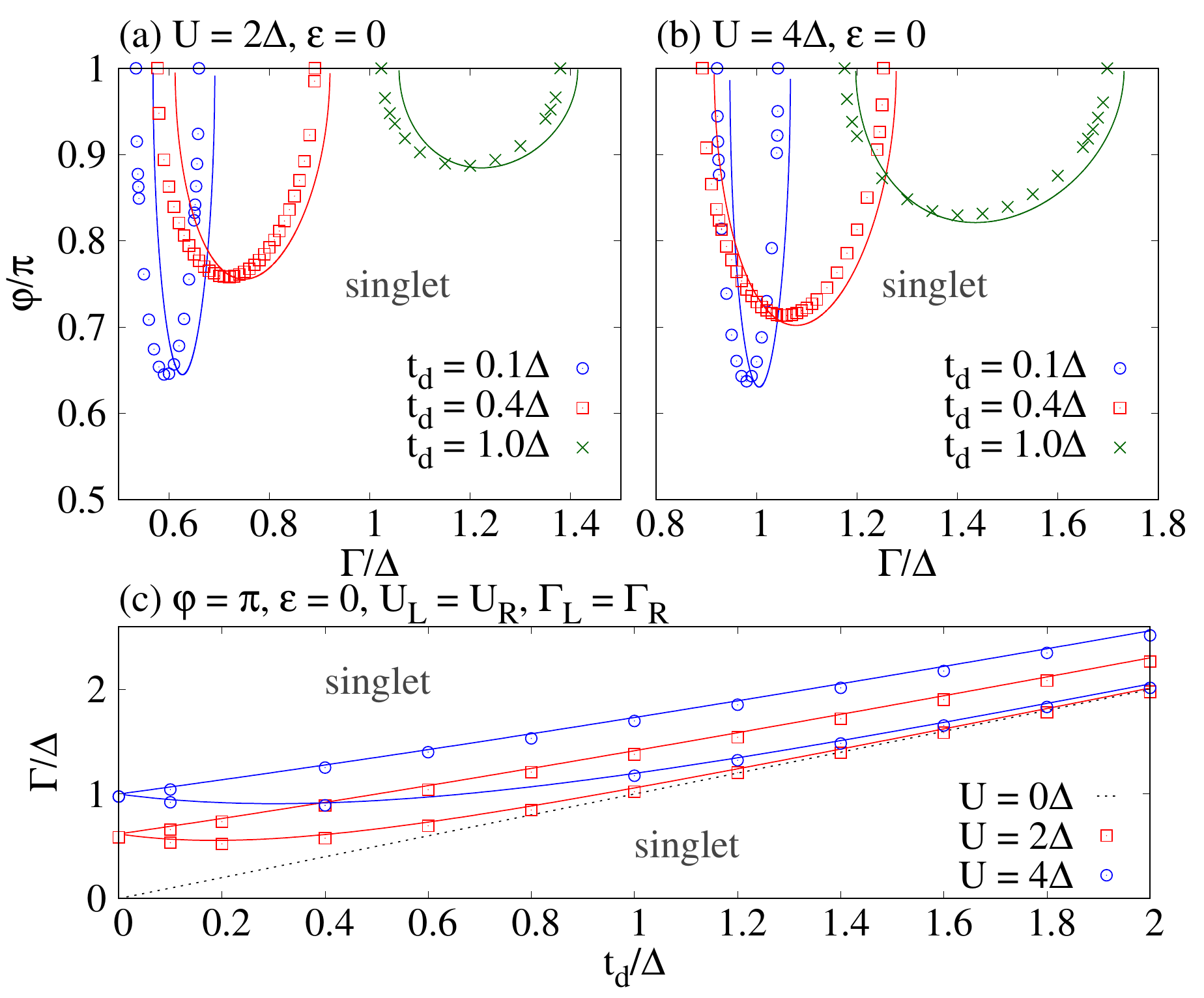}
\caption{
	[(a) and (b)]
	Phase diagrams for three selected values of $t_d$ for symmetric [$\Gamma_L=\Gamma_R=\Gamma$, $U_L=U_R=U=2\Delta$ (a) and $U=4\Delta$ (b)] half-filled case in $\varphi-\Gamma$ plane. The doublet ground states are enclosed by the semi-elliptic phase boundaries while singlets form outside this region. 
	(c) Phase diagrams in $\Gamma-t_d$ plane at $\varphi=\pi$ for $U=2$ and $4\Delta$. The dotted black line $\Gamma=t_d$ marks the GAL boundary between two types of singlet phases at $U=0$. For details see the discussion in the text. In all panels symbols represent NRG results and lines of corresponding colors are GAL predictions.
}
\label{fig:Phi_PhD}

\end{figure}

The suppression of the doublet ground state at half-filling is well demonstrated in the $U$-$\Gamma$ plane as shown in Figs.~\ref{fig:UG_PhD}(a) and~\ref{fig:UG_PhD}(c). Clearly, increasing $t_d$ makes the pockets of doublet ground state smaller and pushes them toward larger dot asymmetries. However, this should not prohibit observation of QPTs in experiments, even at half-filling, since some realizations, e.g., scanning tunneling spectroscopy setups with superconducting tip, may involve a large coupling asymmetry. In addition, although increasing $t_d$ suppresses the doubled phase, it can survive even for hopping terms comparable with the superconducting gap as it is shown in Figs.~\ref{fig:UG_PhD}(b) and~\ref{fig:UG_PhD}(d). Again, all this can be deduced from the inexpensive GAL analysis, which is in good quantitative agreement with the NRG results. 

Allowing for tunable phase difference $\varphi$, as possible in some SQUID-based experiments, significantly enlarges the parametric space. Scanning the multi-parameter phase boundaries with NRG then becomes even more tedious as it requires additional numerical resources. Fortunately, the GAL model can be of help here as well. We illustrate this in Figs.~\ref{fig:Phi_PhD}(a) and~\ref{fig:Phi_PhD}(b) in the $\varphi-\Gamma$ plane for two values of $U$ and various values of $t_d$. Generally, the GAL and NRG results remain in solid agreement and show that a QPT can be observed at half filling even for otherwise perfectly symmetric dots if $\varphi$ is large enough.

%----------------fig-6----------------
\begin{figure}[tbh]
\includegraphics[width=1.0\columnwidth]{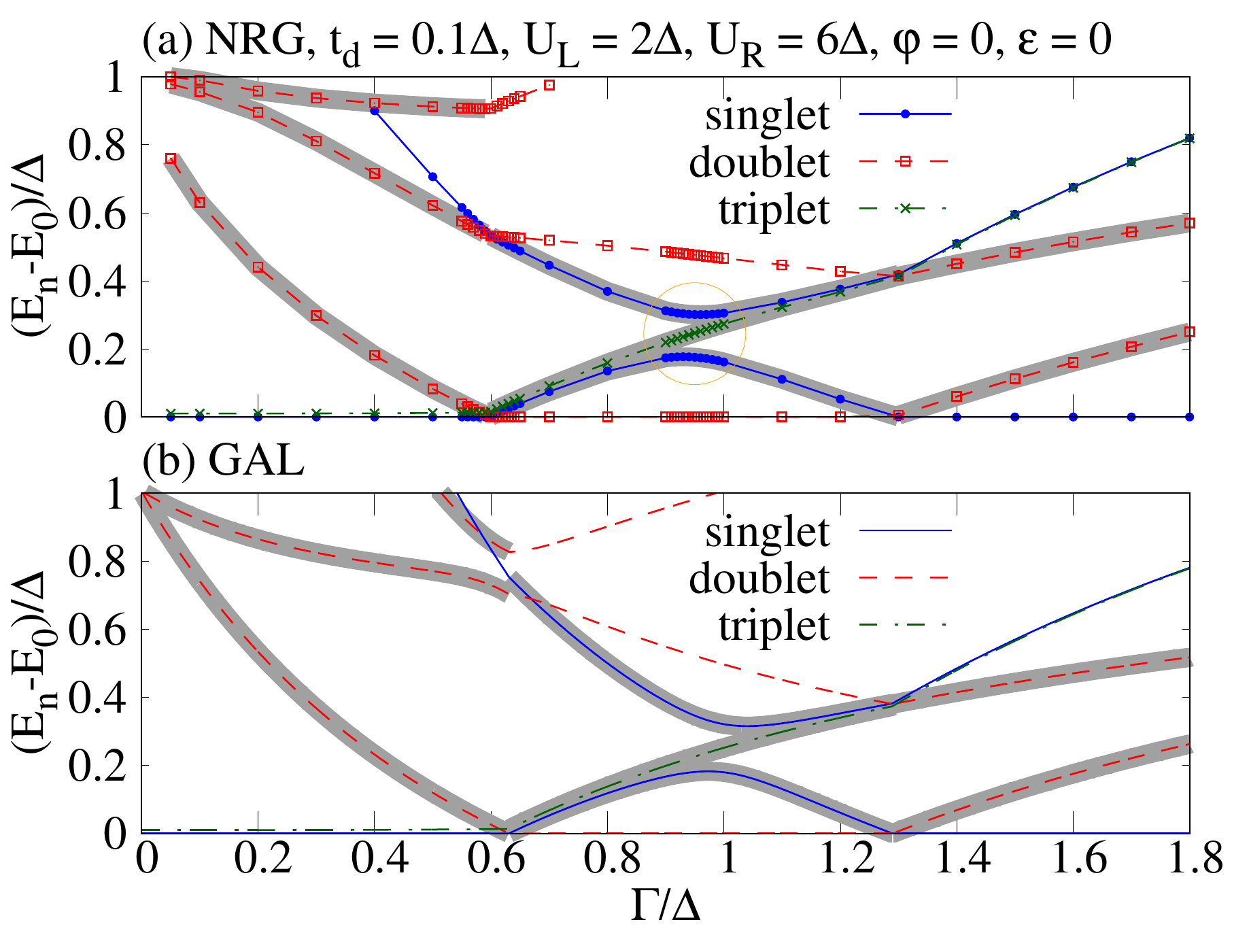}
\caption{
	Example of a complicated subgap state dependence on $\Gamma$ for asymmetric dots at half-filling. (a) depicts the NRG results and panel (b) shows the GAL predictions. There are three different singlet states (blue solid lines and circles), two doublet states (red dashed lines and squares) and one triplet state (green dot-dashed lines and crosses). Note that the differences between the energies of excited states and the ground state energy equals the absolute values of ABS energies if the single particle transition between the states is allowed. The energies of ABS are underscored by gray stripes. The orange circle in (a) marks the avoided crossing between singlet I and single II type of excited states.}
\label{Fig:absG}
\end{figure}

In more detail, Figs.~\ref{fig:Phi_PhD}(a) and~\ref{fig:Phi_PhD}(b) show pockets of doublet phase near $\varphi=\pi$ which have been so-far reported only away from half-filling~\cite{Zitko2010PRL}. Their position and size is strongly influenced by $t_d$. As it increases, the doublet phase is suppressed toward higher values of $\varphi$. Nevertheless, at $\varphi=\pi$ the width of the doublet-phase pocket is relatively stable and as such survives even for $t_d > \Delta$. This is shown in panel (c), where the doublet region is sandwiched between two singlet phases.

Analyzing the GAL model, we can back up these numerical findings analytically. Comparing the eigenenergies in Eqs.~\eqref{eq:EVs} and~\eqref{eq:EVd} from Appendix~\ref{Sec:AppGALH} we get two critical values of $t_d$. Namely:
\begin{eqnarray}
t_d^{c_1}&=& \Gamma-\frac{U}{2(1+\Gamma/\Delta)},
\nonumber \\
t_d^{c_2}&=&\frac{\Gamma}{3} + \sqrt{\left(\frac{2\Gamma}{3}\right)^2-\frac{U^2}{12(1+\Gamma/\Delta)^2}}.
\label{eq:GALphi1}
\end{eqnarray}
They delimit the doublet-phase region at $\varphi=\pi$ for finite $U$. For the non-interacting case ($U=0$), $t^{c_1}_d=t^{c_2}_d=\Gamma$ [black dashed line in Fig.~\ref{fig:Phi_PhD}(c)] the two phase boundaries collapse to a single boundary between the two types of singlet phases with doublet ground state completely eliminated from the phase diagram. Because $\varphi=\pi$ ensures the widest doublet phase pocket, the formulas~\eqref{eq:GALphi1} also limit the parameters $t_d$, $U$ and $\Gamma$ or their combinations for which the doublet phase can be observed for otherwise symmetric dots. We discuss this in more detail in Appendix~\ref{Sec:AppDP}.

Note that to have a finite phase difference $\varphi$ the system has to have two leads. In the limit $t_d\rightarrow 0$ we practically split the double dot into two independent systems. Each has just one electrode. Therefore, for $t_d\rightarrow 0$ we are always combining single dots with $\varphi=0$. Consequently, the $\Gamma$ which separates the singlet states of type I and II at $t_d\rightarrow 0$ in Fig.~\ref{fig:Phi_PhD}(c) can be read out from the phase diagrams in Fig.~\ref{fig:UU_PhD}(a) for any $U$.

%--------------fig-7-------------
\begin{figure}[tbh]
\includegraphics[width=1.0\columnwidth]{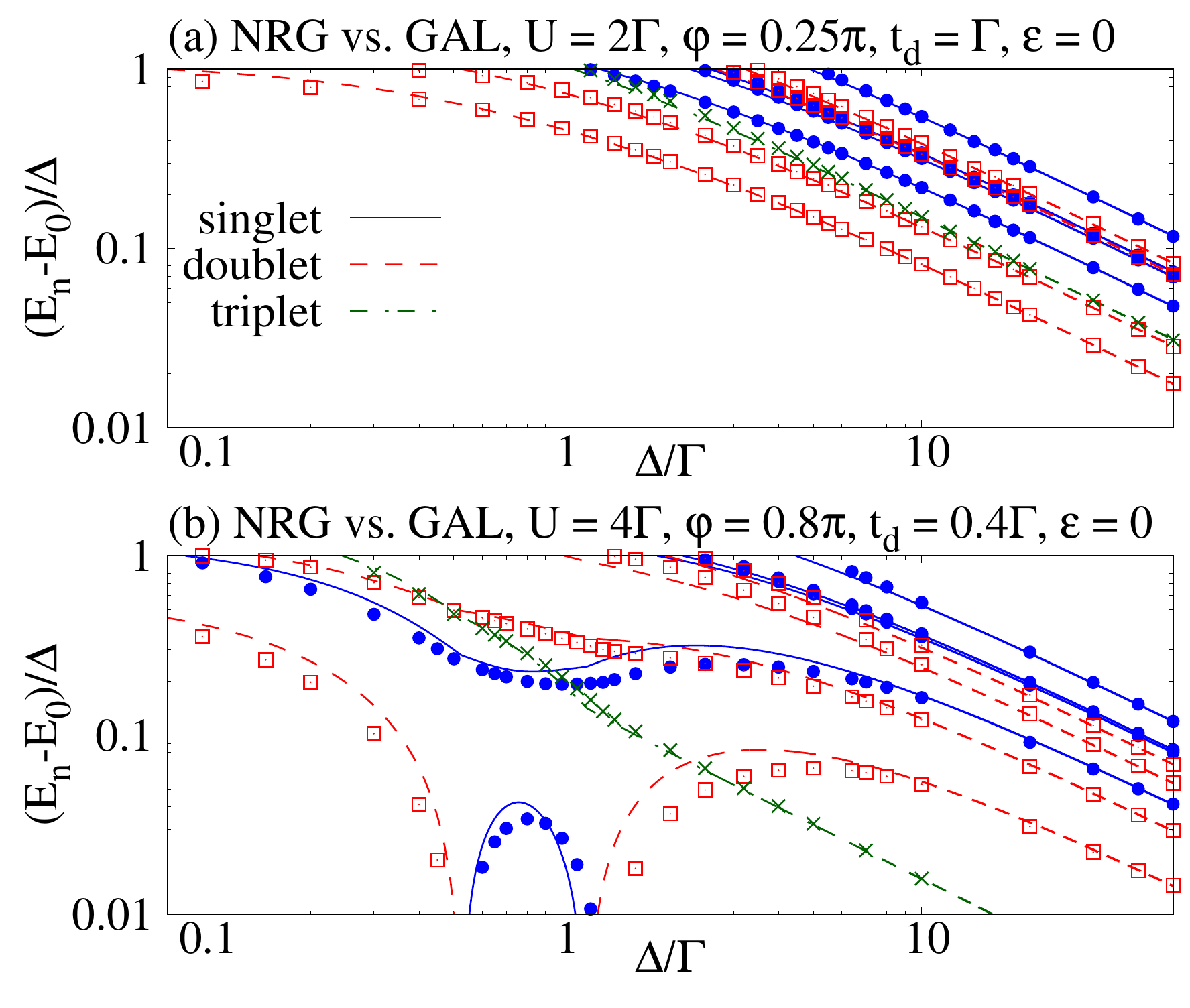}
	\caption{Direct comparison of NRG (symbols) and GAL (lines) results for subgap many-body states as functions of
	$\Delta$ calculated for two sets of parameters. Note the logarithmic scales.
    There are four different singlet states (blue solid lines and circles) in panel (a) and five in panel (b),  four doublet states (red dashed lines and squares) and one triplet state (green dot-dashed lines and crosses) in both panels.}
\label{Fig:absDeltaDep}
\end{figure}

%%%%%%%%%%%%%%%%%%%%%%%%%%%%%%%%%%%%%%%%%%%%%%%%%%%%%%%%%%%
\subsection{Subgap states at half-filling}
\label{sec:ABS}

One of the main benefits of the GAL model for single QD was the quantitative correction of the ABS profile of the original AL theory. For SDQDs, the same is demonstrated in Fig.~\ref{Fig:absG} with subgap  many body states [$(E_n - E_0) < \Delta$] calculated for parameters $U_L=2\Delta$, $U_R=6\Delta$, $t_d=0.1\Delta$ and $\varphi=0$ via NRG [panel (a)] and GAL model [panel (b)].
Note that despite its simplicity the GAL approximation captures correctly all of the main features
of the subgap spectrum as discussed in more detail below. In addition, with the exception of
the region of small $\Gamma$ the energies of the subgap states are in a very good quantitative agreement with the NRG as well. In Appendix~\ref{Sec:AppSS}, we directly compare the energies of the NRG and GAL subgap states as functions of $\varphi$ for different parameter settings. Here, we instead focus on the qualitative aspects. 
 
In this regard, we stress that the agreement between NRG and GAL is impressive considering the complexity of the subgap states. For example, the behavior of the ground state and the first excited state in Fig.~\ref{Fig:absG} reflects the
reappearance of the singlet phase separated by doublet phase shown in Fig.~\ref{fig:UG_PhD}.  
As discussed for $t_d\rightarrow 0$, the two singlet regimes have different origins.
The one at small $\Gamma$ has a type II singlet ground state. The first excited state in this region is, therefore, a triplet state (dot-dashed green line) emerging from the singlet-triplet splitting. Because the splitting energy is of the order of $t^2_d/(U_L+U_R)$ and $t_d=0.1\Delta$, the
triplet state closely ground state. 
The second excited state in this region is a doublet state (dashed red line) which at the QPT point $\Gamma\approx0.6\Delta$ becomes the ground state.
In the doublet phase the first excited state is singlet (solid blue line), which 
becomes again the ground state at the second QPT point at $\Gamma\approx1.3\Delta$. 
However, following the discussion for the $t_d\rightarrow 0$ case, the ground state of this second singlet phase is the type I singlet. Therefore, it is not accompanied by a triplet state and the first as well as the second excited states are doublets (red dashed lines). Only above them the triplet state closely follows a type II singlet. An avoided crossing of type I and type II singlets can be seen in the central part of the doublet phase marked by orange circle in panel (a). The GAL model in panel (b) correctly captures all of these details. We would like to stress here that the excitations singlet-to-triplet as well as singlet-to-singlet violate the $\Delta S_z = 1/2$ selection rule and, therefore, will not be visible in the one-electron spectral function, i.e., not all of the excited states will contribute to ABS~\cite{Zitko2015dd}. The energies of the allowed transitions, i.e., ABS, are underscored with gray stripes in Fig.~\ref{Fig:absG}.

Considering that the GAL model is based on AL theory with scaled parameters that reintroduce the finite gap, it is important to check how the GAL model reacts to evolving $\Delta$. Therefore, in Fig.~\ref{Fig:absDeltaDep} we selected two very distinct sets of model parameters and tested the GAL model against NRG solutions in a wide range of $\Delta/\Gamma$ values. Their agreement increases with increasing $\Delta/\Gamma$ as expected for AL theory. Nevertheless, even for small $\Delta$, i.e., several times smaller than any other energy parameter, the GAL model gives surprisingly good predictions for the positions of subgap states for a tiny fraction of the NRG computational costs.
        
%%%%%%%%%%%%%%%%%%%%%%%%%%%%%%%%%%%%%%%%%%%%%%%%%%%%%%%%%%%
\subsection{Josephson current at half-filling}
\label{sec:JC}

%--------------fig-8-------------
\begin{figure}[tbh]
	\includegraphics[width=1.0\columnwidth]{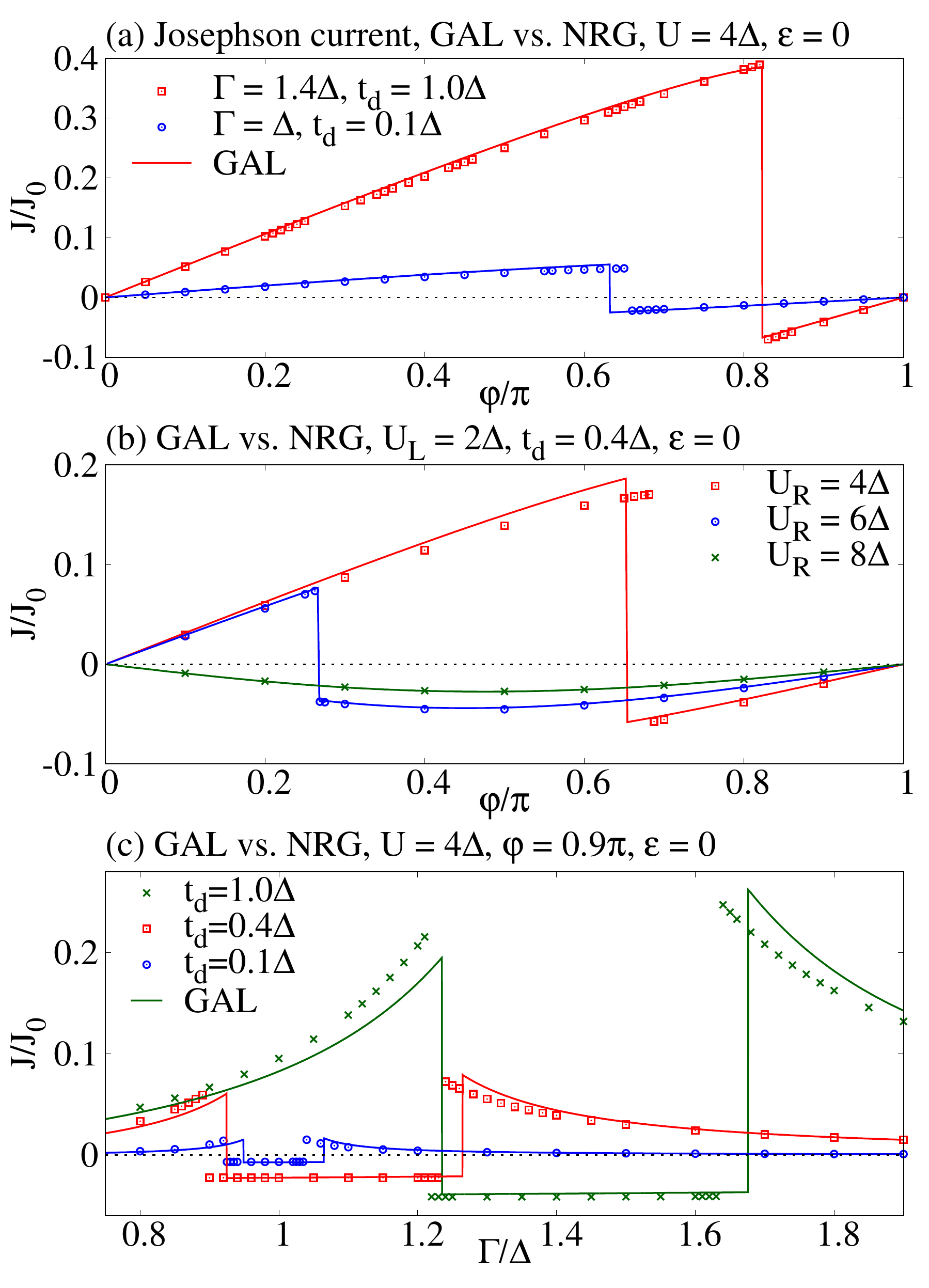}
	\caption{
		Direct comparison of the Josephson current calculated for various parameters via NRG (symbols) and GAL (solid lines), where $J_0=e\Delta/\hbar$. (a) and (b) show the $\varphi$ dependence. In (a) the changes of the sign coincide with the respective singlet-doublet phase boundaries shown in Fig.~\ref{fig:Phi_PhD}(a) for symmetric case. (b) represents an asymmetric case.  (c) shows the $\Gamma$ dependence for $\varphi=0.9\pi$, therefore, the change of current sign is close to the respective phase boundaries in Fig.~\ref{fig:Phi_PhD}(c) where we used $\varphi=\pi$.}
	\label{fig:SC2d}
\end{figure}

The GAL model captures correctly also the Josephson current. We illustrate this in 
Fig.~\ref{fig:SC2d}, where we compare the GAL predictions with the NRG results for various parameters as a function of $\varphi$ [panels (a) and (b)] and $\Gamma$ [panel (c)]. In contrast to its single dot version, the GAL model for the double-dot system correctly predicts the current in both phases. Up to a small shift in the predicted position of the phase transition point, the GAL Josephson current follows the NRG results. It is sensitive to $U$ in the singlet phase and non-zero in the doublet phase. There was no necessity to introduce any band corrections, as the
doublet ground state energy is phase-dependent for SDQD. This is true for both symmetric (a),(c) and non-symmetric dots (b). It indicates that, in contrast to the single dot case, the Josephson current for SDQD in the doublet phase at half-filling is predominantly carried by the ABS and not the incoherent band states.

%%%%%%%%%%%%%%%%%%%%%%%%%%%%%%%%%%%%%%%%%%%%%%%%%%%%%%%%%%%
\subsection{Away from half-filling: MGAL phase boundary scan}
\label{sec:Prel}

\begin{figure}
	\includegraphics[width=1.0\columnwidth]{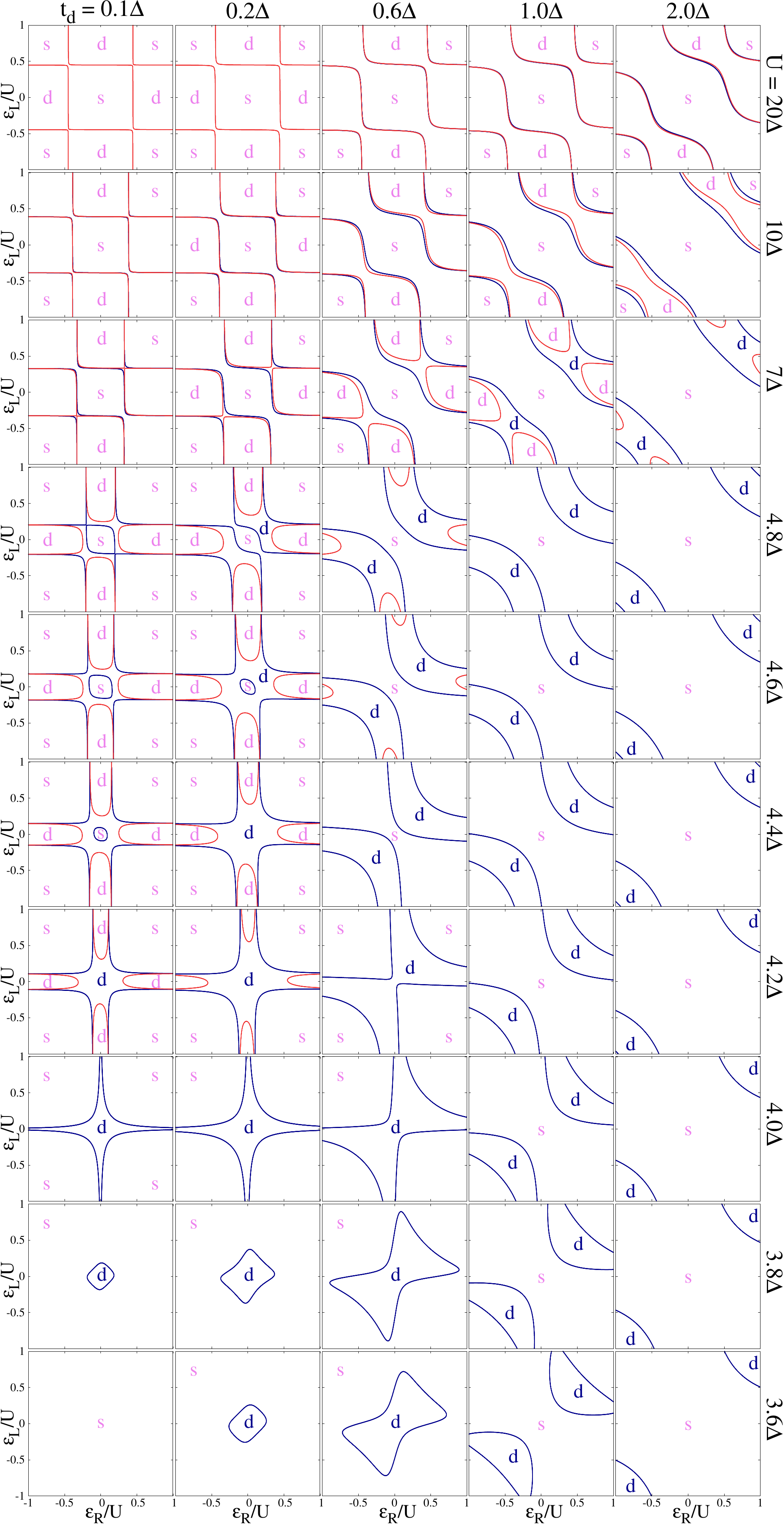}
	\caption{Evolution of the phase boundaries for the symmetric case with $U_L=U_R=U$ and $\Gamma_R=\Gamma_L=\Gamma=\Delta$, $\varphi=0$ (red lines) and $\varphi=\pi$ (blue lines) respectively. Phase diagrams are plotted in the $\varepsilon_L/U$-$\varepsilon_R/U$ planes (from $-1$ to $1$ on each axis) for different combinations of $t_d$ and $U$. The rows show (from bottom up) the results for $U=3.6$, $3.8$, $4.0$, $4.2$, $4.4$, $4.6$, $4.8$, $7$, $10$, $20\Delta$. The columns show from left to right the results for $t_d=0.1$, $0.2$, $0.6$, $1$, $2\Delta$. Particular regions are marked by s for singlet and d for doublet. The violet letters signal that both $\varphi=0$ and $\varphi=\pi$ have the same type of ground state in the marked region. On the other hand, blue d means that the ground state is doublet only for the $\varphi=\pi$ case.}
	\label{Fig:phd_Map}
\end{figure}

The half-filled case discussed so far plays a crucial role in the analysis of any structure of QDs coupled to a superconductor. Nevertheless, the filling of each dot in the SDQD system can be controlled in some experiments via respective gate voltages~\cite{Saldana2018, Saldana2022,Saldana2020}. Combining the electrostatic control of energy levels $\epsilon_L$ and/or $\epsilon_R$ of each dot with the differential conductance measurements ensures a high degree of tunability. However, it also enlarges the available parameter space and, therefore, pushes its theoretical analysis via NRG or QMC to the limits of their practical usability.

To reproduce the experimental charge stability diagrams one can analyze the SCIAM in the $\varepsilon_L/U_L - \varepsilon_R/U_R$ plane since in the limit of weakly coupled dots $\langle n_i \rangle \approx 1 + \varepsilon_i/U_i$~\cite{Rasmussen2018,Saldana2018}. However, to produce maps with a needed resolution is extremely costly when NRG and/or QMC calculations are employed. This gets even more complicated when some of the model parameters are unknown, or not known with a sufficient precision.  
On the other hand, the MGAL model can be easily utilized for a detailed preliminary scan of the parameter space on a standard PC. This led us to the discovery of some interesting regimes.

Here we focus on a symmetric scenario with $U_L=U_R=U$ and $\Gamma_R=\Gamma_L=\Gamma=\Delta$. In Fig.~\ref{Fig:phd_Map} we show MGAL phase boundaries in the $\varepsilon_L/U_L - \varepsilon_R/U_R$ plane for $\varphi=0$ (red lines) and $\varphi=\pi$ (blue lines) at varying values of $t_d$ and $U$. The figure captures a complicated evolution of the phase diagram. Starting at $U\gg t_d$, e.g., $t_d=0.1\Delta$ and $U=20\Delta$ (upper left corner in Fig.~\ref{Fig:phd_Map}), we identify a rather trivial phase diagram that resembles a disconnected SDQD system ($t_d\rightarrow0$) with an emerging checkerboard pattern of singlet and doublet phases (with the singlet one in the center). Only very small bending is observed at what would be the quadruple degeneracy points for $t_d=0$ due to $t_d/U \rightarrow 0$. In this limit the dots are only weakly linked, therefore, the $\varphi=\pi$ and $\varphi=0$ cases are indistinguishable. Consequently, the phase diagrams are insensitive to the presence of the phase bias.

Keeping a constant $t_d=0.1\Delta$ we can observe how decreasing $U$ shapes the phase boundaries. Initially, both $\varphi=0$ and $\varphi=\pi$, evolve indistinguishably as increased $t_d/U$ ratio induces stronger bending of the parity transition lines around the (almost) quadruply degenerated points.
However, the doublet phases disconnect for $\varphi=0$ at $U\approx5\Delta$ and form four isolated regions. These are pushed by the decreasing $U$ to higher values of $\varepsilon_j/U$ until they completely vanish from the plotted regions at $U\approx 4 \Delta$ leaving only a singlet ground state in the plotted phase space.
The evolution for $\varphi=\pi$ is different.
Instead of splitting, the doublet phases merge into one region, which at $U\approx4.6\Delta$ leads to a formation of a closed pocket of the singlet phase around the half-filled point $\varepsilon_L=\varepsilon_R=0$. With further decreasing of $U$
this central singlet phase pocket shrinks until it completely vanishes for $U\approx4.2\Delta$. Simultaneously, the doublet phase shrinks as well. First it evolves into an elongated four-pointed star-like pattern with no phase transitions appearing along the $\varepsilon_L=0$ and $\varepsilon_R=0$ lines. Next, upon further decreasing of $U$, the branches of the star-like pattern connect. Therefore, in this region, the doublet phase forms a pocket in the center of the plane, e.g., for $U \approx3.8\Delta$.

If we now fix $U \approx3.8\Delta$ and let $t_d$ increase, we see how this pocket again grows and eventually splits into two independent doublet regions. Going back to the strong interaction, a similar evolution in $t_d$ forms bended stripes of alternating singlet and doublet phases. 

These patterns are the most common outcomes in stability diagrams measured experimentally, for an example see Ref.~\cite{Saldana2018}. Both the $\varphi=0$ and $\varphi=\pi$ phase boundaries show such phase orderings for $U \gtrsim 10\Delta$. However, the $\varphi=0$ doublet stripes are less stable. With decreasing $U$ and/or increasing $t_d$ they disconnect as seen for example for $U=7\Delta$ and $t_d=0.6\Delta$. The split parts contract upon decreasing $U$ and $t_d$, which consequently leaves only a singlet phase present even for large $U$.

For $\varphi=\pi$ the stripe pattern persists even at moderate parameters such as $t_d \approx 0.6 \Delta$ and $U \gtrsim 4.4\Delta$. Nevertheless, for smaller $U$ these stripes merge either into already discussed star-like shapes for small $t_d$ or isolated doublet pockets for moderate $t_d$.

Both $\varphi=0$ and $\varphi=\pi$ eventually collapse into trivial singlet regimes if $U$ or/and $t_d$ are small enough, where no QPT exists. Nevertheless, it is important to stress two aspects here. First, the doublet phases for $\varphi=0$ (actually, any $\varphi<\pi$) are encapsulated within doublet phases for $\varphi=\pi$. Second, for $\varphi=\pi$ the doublet phase survives to much lower values of $U$ and $t_d$ than for $\varphi=0$. In Fig.~\ref{Fig:phd_Map} the trivial singlet outcome appears for all panels with $\varphi=0$ and $U<4.2\Delta$. Yet only a single panel shows such an outcome for $\varphi=\pi$ ($U=3.6\Delta$ and $t_d=0.1\Delta$).

\begin{figure}[tbh]
	\includegraphics[width=1.0\columnwidth]{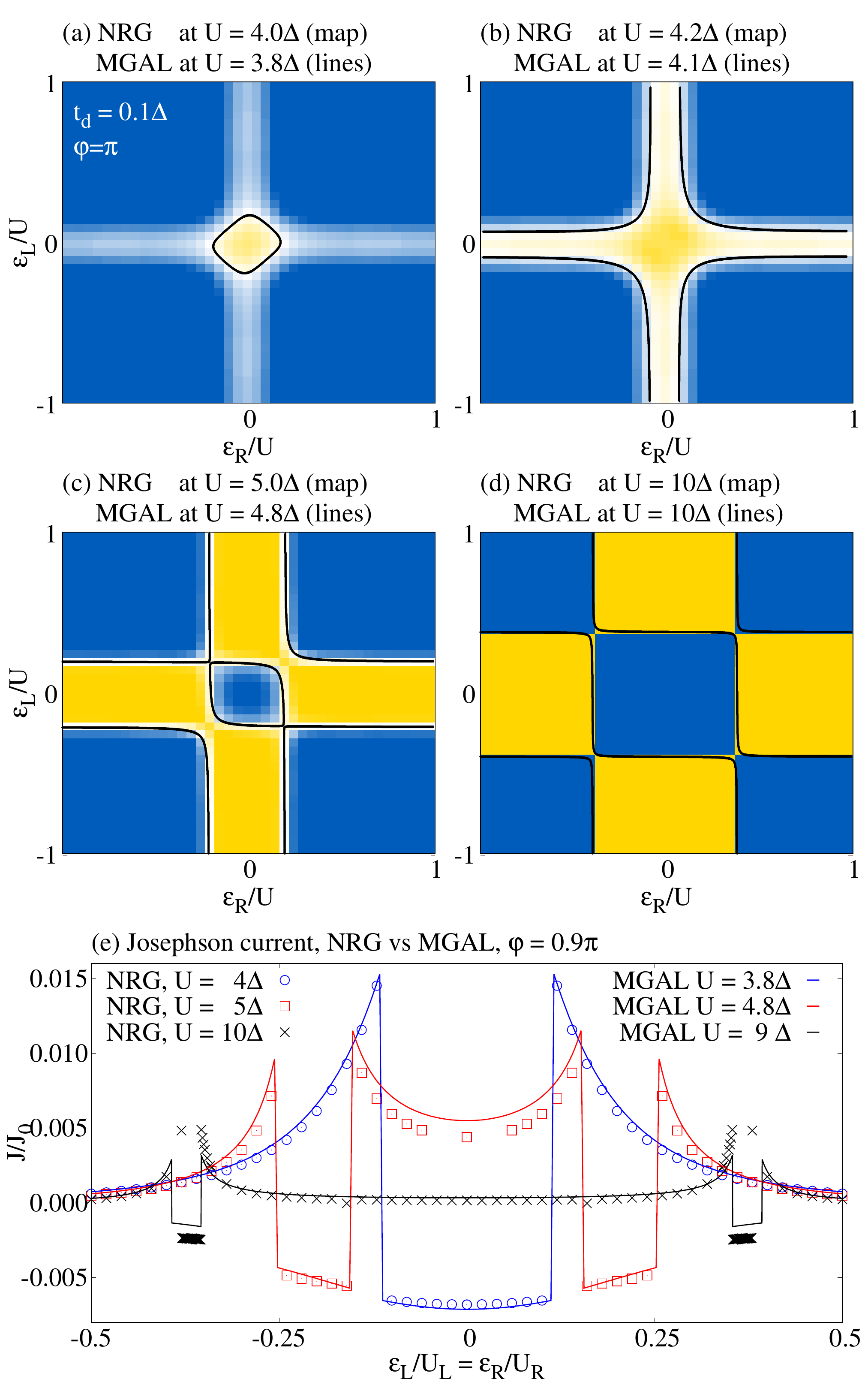}
	\caption{ Evolution of phase (stability) diagrams in the $\varepsilon_L/U_L$ - $\varepsilon_R/U_R$ plane at $\Gamma=\Delta$, $t_d=0.1\Delta$ and $\varphi=\pi$ upon increasing $U$. Color-maps show the difference between the lowest singlet and lowest doublet eigenenergies calculated via NRG. The blue color denotes the stability region of a singlet phase and the yellow color of a doublet phase. The black curves show the MGAL phase boundaries. (d) The Josephson current along the diagonal $\varepsilon_L=\varepsilon_R$ calculated with NRG (symbols) and MGAL (solid lines).  Note that MGAL employs slightly shifted Coulomb energies with respect to NRG. For explanation of the shift see the main text. 
	}
	\label{fig:mapU}
\end{figure}

%%%%%%%%%%%%%%%%%%%%%%%%%%%%%%%%%%%%%%%%%%%%%%%%%%%%%%%%%%%
\subsection{Away from half-filling: comparison with NRG \label{sec:AHF}}

The MGAL scan of the parameter space predicts the existence of regions where a small change in $U$ or $t_d$ leads to a dramatic evolution of the $\varepsilon_L/U - \varepsilon_R/U$ phase diagrams. Considering that even at half-filling the GAL phase boundaries are not perfectly aligned with NRG points, one can expect in this region a mismatch between MGAL and NRG. This is indeed the case. Nevertheless, this issue can be often solved by a small shift of selected MGAL parameters, as we discuss in this section. Its main purpose is to test the MGAL predictions against the NRG results and to establish the validity bounds of the MGAL approximation away from half-filling.

We first explore the case of $t_d=0.1\Delta$ with $\Gamma_L=\Gamma_R=\Gamma=\Delta$, at varying $U=U_L=U_R$.  
We have selected $\varphi=\pi$ because this case shows the most complex and most stable (with respect to parameter change) structures in the MGAL analysis. Both NRG and MGAL predict a trivial singlet phase for $U \lesssim 3.8\Delta$ without any doublet phases. We, therefore, omit this regime. The results for four higher values of $U$, representing different phase diagram regimes, are shown in Fig.~\ref{fig:mapU}. 
At $U = 4 \Delta$ [panel (a)] the NRG confirms a small pocket of doublet phase in the center of the diagram as predicted by MGAL.
Here the color-map was obtained by NRG and it shows a difference between the lowest singlet and lowest doublet eigenenergies. 
Therefore, the blue color signals a stable singlet phase (negative values) and yellow the doublet phase (positive values). 
The black line marks the corresponding phase boundary provided by MGAL. It coincides with the (white) transition area in the NRG map. 

As we tune $U$ up to $\approx 4.2\Delta$ the NRG result confirms a significant change of the phase diagram as predicted by MGAL. Regions of the doublet ground state elongate along the $\varepsilon_L=0$ and $\varepsilon_R=0$ axes. The phase diagram becomes star-shaped with doublet ground state in the center and with no signs of QPTs along the $\varepsilon_L=0$ and $\varepsilon_R=0$ lines as shown in Fig.~\ref{fig:mapU}(b).

When $U$ is further increased the expected central singlet island emerges in the NRG data as illustrated by panel (c) for $U=5\Delta$. 
Both MGAL and NRG show wide doublet branches of the former star-like pattern stretching along the $\varepsilon_L=0$ and $\varepsilon_R=0$ lines. 
 
Finally, for strong $U$ the overall pattern indeed resembles a regular rectangular checkerboard consisting of singlet and doublet regions. Initially, for sufficiently high ratio of $t_d/U$, there is a clear bending of the phase boundaries at the parity transition lines. However, this is strongly suppressed as $t_d/U\rightarrow0$ as shown in Fig.~\ref{fig:mapU}(d) for $U=10\Delta$. 

Note that in Figs.~\ref{fig:mapU}(a)-(c) we have used slightly smaller (within $5\%$) values of $U$ for MGAL than for the NRG calculations. As already discussed, a small variation of parameters $U$ and $t_d$ leads in the discussed region to a qualitative change of the phase diagram. This easily leads to a situation where MGAL and NRG phase boundaries calculated for exactly the same $U$ and $t_d$ predict a different type of phase diagrams.
However, a small constant shift of $U$ (or $t_d$) solves this problem. After such trivial reparametrization the resulting phase boundaries from MGAL are in agreement with NRG in the whole investigated $\varepsilon_L/U - \varepsilon_R/U$ plane. 

Moreover, this small correction also leads to very good agreement between the Josephson current calculated with NRG and the MGAL model. We show this in panel (e) where the current is plotted as a function of $\varepsilon_L/U = \varepsilon_R/U$, i.e., it follows the diagonal in the plotted phase diagrams (a),(b) and (c). Although, we use $\varphi=0.9\pi$ instead of $\pi$, because at $\varphi=\pi$ there is no supercurrent. Clearly, even for moderate $U$, represented by $U=4\Delta$ (blue) and $U=5\Delta$ (red) in panel (e) alike some experimental realizations~\cite{Saldana2020}, there is a good quantitative agreement between NRG and MGAL. This, however, changes when we push $U$ further into the strong interaction limit.
 
Seemingly, no parameter adjustments are needed for $U \gg \Delta \gg t_d$, as the MGAL and NRG phase boundary are nicely aligned in Fig.~\ref{fig:mapU}(d) for the same $U=10\Delta$. However, this is related to the checkerboard pattern which is stabilized at $U=10\Delta$. Besides a small bending of the corners of the central square, there are not enough details to distinguish diagrams with similar $U$ (ot $t_d$) in this regime. MGAL can, therefore, predict the phase boundaries with great accuracy. Nevertheless, the MGAL Josephson current in panel (e) shows for strong $U=10\Delta$ a much larger quantitative difference from the NRG results than for intermediate $U$. The position of the phase transition can be tuned by adjusting $U=9\Delta$. However, in the vicinity of the QPT the amplitude of the current differs significantly from the NRG result. Here the NRG Josephson current can be more than twice the MGAL Josephson current. Although this is still not a bad result for an effective model, it points to the limitations of MGAL in the strongly interacting regime.

\begin{figure}[tbh]
	\includegraphics[width=1.0\columnwidth]{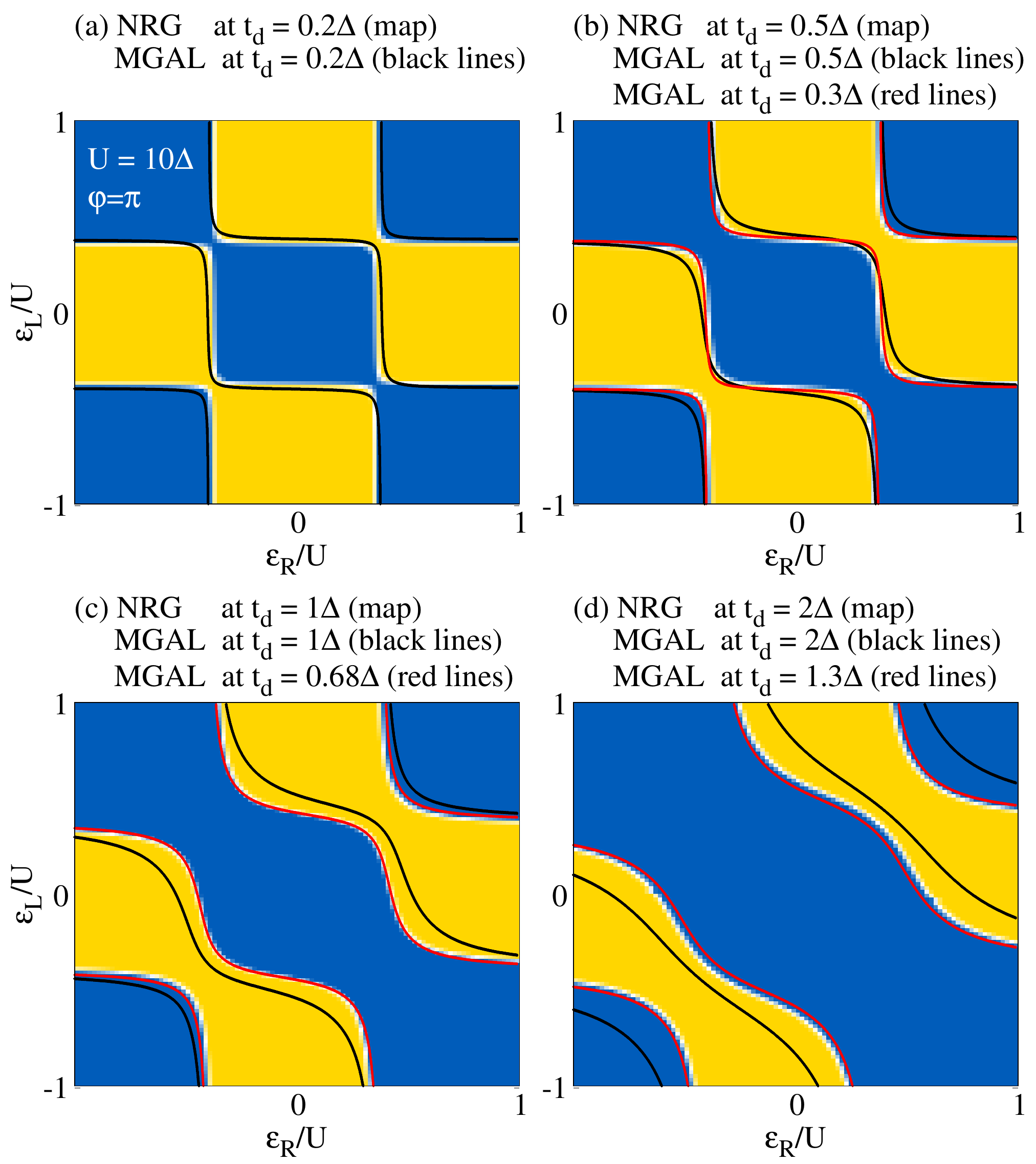}
	\caption{Evolution of phase (stability) diagrams in the $\varepsilon_L/U_L$ - $\varepsilon_R/U_R$ plane at $\Gamma=\Delta$, $U=10\Delta$ (NRG) and $\varphi=\pi$ upon increasing $t_d$. Color-maps show the difference between the lowest singlet and lowest doublet eigenenergies calculated via NRG. The blue color signals the stability region of a singlet phase and the yellow color of a doublet phase. In the vicinity of the phase boundaries the color changes to white. The black curves show the MGAL phase boundaries calculated for $U=10\Delta$. The red curves in (b)-(d) show MGAL results for shifted values of $U$ for which MGAL results give a better agreement with NRG data.}
	\label{fig10}
\end{figure}
Considering that in experiments one can have $U\gg \Delta$ and simultaneously $t_d \approx \Delta$~\cite{Saldana2018}, it is worth looking into how the diagram in Fig.~\ref{fig:mapU}(d) evolves with increasing $t_d$.
We illustrate this in Fig.~\ref{fig10} where we show diagrams at $U=10\Delta$, $\Gamma=\Delta$ and $\varphi=\pi$ but now we gradually increase $t_d$ from $0.2\Delta$ (a) to $2\Delta$ (d). In general, the increasing intra-dot hopping causes bending of the phase boundaries. Consequently, the checkerboard pattern evolves into a diagonal stripe-like phase diagram. This is supported by both NRG (color map) and MGAL (black lines). However, a direct quantitative comparison between NRG and MGAL reveals that their phase boundaries coincide only for small $t_d$ [e.g., $t_d=0.2\Delta$ in panel (a)].  To get a quantitative agreement for higher $t_d$ we have to adjust the MGAL parameters. This time we adapt $t_d$ instead of $U$. Unfortunately, a much larger shift is needed here ($\approx30\%$). Nevertheless, once again a constant shift of $t_d$ is sufficient to reproduce NRG results in the whole $\varepsilon_L/U -\varepsilon_R/U$ plane. We show this in Fig.~\ref{fig10}(b)-(d) where the red lines are the MGAL phase boundaries calculated with shifted $t_d$ (see the description above the panels for particular values) keeping all other parameters the same as in the NRG solutions. The great agreement between red-lined MGAL boundaries and NRG illustrates the strength of the generalized AL approach. On the other hand, the large parameter shifts also clearly show the limitations of the MGAL model in this particular regime. 

\begin{figure}[tbh]
	\includegraphics[width=1.0\columnwidth]{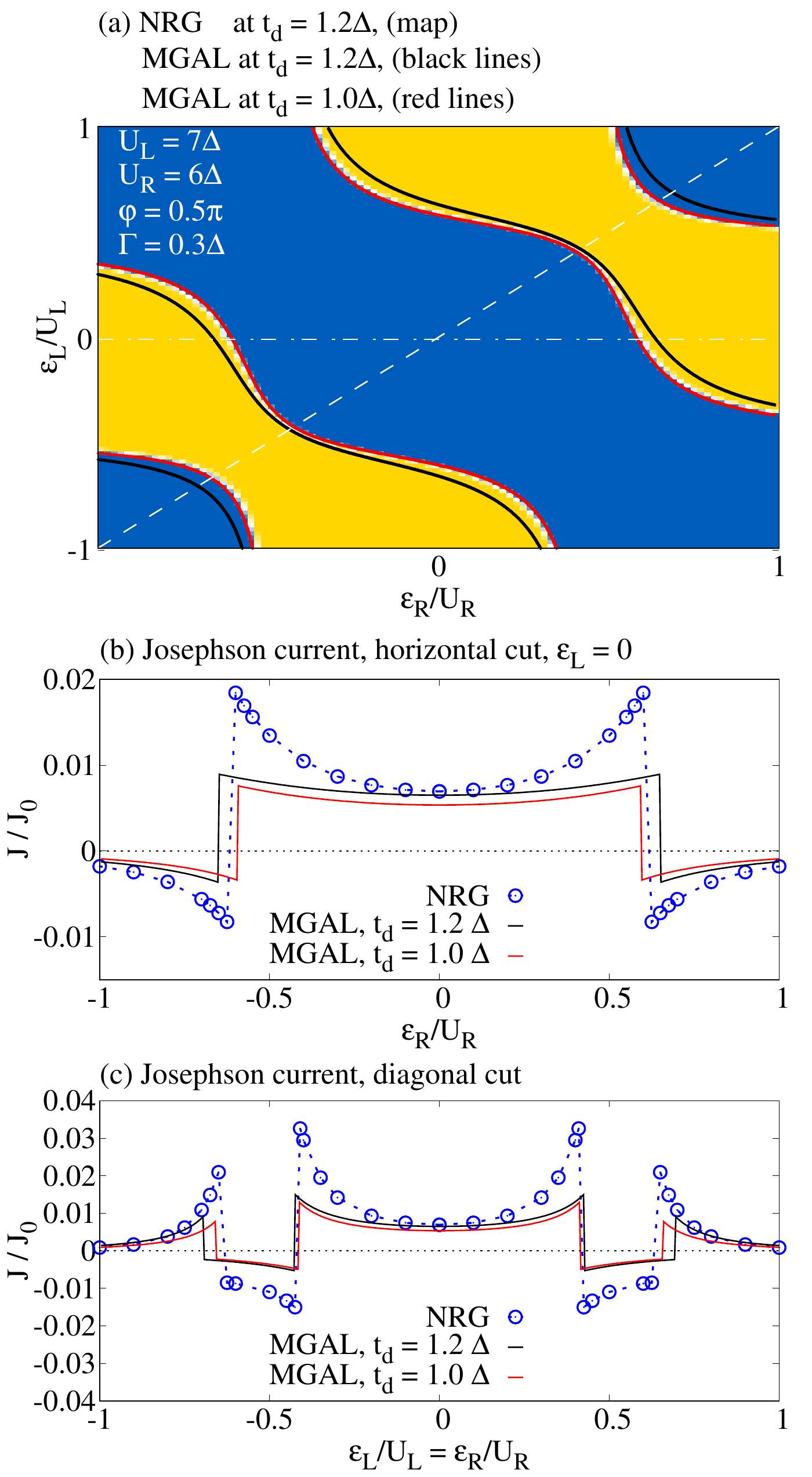}
	\caption{Analysis of an experimental setup from Ref.~\cite{Saldana2018}. Here, $U_L=7\Delta$, $U_R=6\Delta$, $t_d=1.2\Delta$, $\Gamma=0.3\Delta$, $\varphi=0.5\pi$. (a) Phase diagram in the  $\varepsilon_L/U_L$ - $\varepsilon_R/U_R$ plane. The color-map shows the difference between the lowest singlet and lowest doublet eigenenergies calculated via NRG. The blue color signals the stability region of a singlet phase and the yellow color of a doublet phase. The black curves show the MGAL phase boundaries for the same parameters as in NRG. The red lines mark MGAL phase boundaries with $t_d=\Delta$. [(b) and (c)] Josephson current for horizontal (b) and diagonal (c) cuts marked by the white dashed lines in panel (a). } 
	\label{fig:exp}
\end{figure}  

Therefore, a question arises as to whether the MGAL model is also applicable to the strongly coupled regime which is often relevant for experiments. To test this we investigate a case for which the parameters had been taken from the experimental setup discussed in Ref.~\cite{Saldana2018}. Namely, we fix $U_L=7\Delta$, $U_R=6\Delta$, $t_d=1.2\Delta$, $\Gamma=0.3\Delta$, $\varphi=0.5\pi$ and focus on changing $\epsilon_j/U_j$. 
We present the NRG and MGAL results in Fig.~\ref{fig:exp}. Panel (a) shows the NRG stability diagram (color map) and the MGAL phase boundaries (black lines). Despite being in the regime of strong coupling $U_j \gg \Gamma$, the agreement is reasonably good taking into account the simplicity of the MGAL model. In addition, a small constant modulation of $t_d$ ($t_d=\Delta$) is sufficient for MGAL to faithfully reproduce the NRG phase boundary as it is shown by the red lines in panel (a). 
In this respect, the MGAL can be indeed useful for the analysis of experiments. However, because of the large $U$ the MGAL model predictions for the Josephson current are much less precise. We show this in panels (b),(c) where the Josephson current is plotted as a function of $\varepsilon_R/U_R$ for $\varepsilon_L=0$ (b) [horizontal cut in panel (a)] and as a function of  $\varepsilon_R/U_R= \varepsilon_L/U_L$ [diagonal cut in panel (c)]. Here the blue circles show the NRG results, black lines the MGAL solution for $t_d=1.2\Delta$ and red one the solution for $t_d=\Delta$. There is a clear discrepancy between NRG and MGAL. Although the correction of the $t_d$ leads to a better location of the QPT points it also makes the magnitude of the Josephson current slightly smaller and, therefore, further away from NRG data. However, here it is important to stress that this disagreement looks bad only when compared to the success of the GAL model in the half-filling or to the MGAL results for intermediate $U$. When compared to other effective models, e.g., ZBW, what is shown in panels (b) and (c) is still a solid result as the currents of other effective models can be off by several orders of magnitude~\cite{Saldana2018}.

%%%%%%%%%%%%%%%%%%%%%%%%%%%%%%%%%%%%%%%%%%%%%%%%%%%%%%%%%%%
\section{Summary \label{sec:summary}}

Effective theories like AL and ZBW are known to capture some of the qualitative properties of the SCIAM model but they fail quantitatively. Here we have introduced a rescaling of the AL theory that overcomes its usual shortcomings. The rescaling is based on the GAL formula which was obtained perturbatively and which is known to correctly capture the phase boundaries for a broad range of parameters. In the case of single dot, the effective GAL model presented here gives not only the correct position of QPT but also a very good quantitative prediction for the position of subgap many-body states and, therefore, ABS. However, its main advantage is that it can be easily generalized to more complex setups which present a significant challenge to NRG or QMC.

We have discussed this in detail for the case of two dots coupled in series. While there is no simple formula for the phase boundaries for the SDQD case, the GAL transformations can be generalized to this case starting from the vanishing intradot coupling limit. At half-filling, the resulting GAL model gives very good quantitative predictions for the position of QPT and subgap states for a broad range of experimentally relevant parameter regimes. Moreover, unlike for the single-dot case, GAL for SDQD gives correct Josephson current in both singlet and doublet phase as confirmed by comparing the GAL predictions with the NRG results. The typical difference between GAL and NRG  was for the most relevant cases within a few percent and, as such, often below the resolution of a typical experiment. The effective GAL model, therefore, allows a fast and reliable analysis of relevant regimes for a tiny fraction of the costs of NRG or QMC.

The GAL model, however, needs some adjustments when used away from half-filling. First of all, the energy levels on the dots have to be modified according to a phenomenological MGAL formula. The MGAL model then gives a solid prediction of the phase boundaries in the $\epsilon_L/U_L-\epsilon_R/U_R$ plane which are crucial for understanding experimental (charge) stability diagrams. Moreover, outside the strong interaction limit ($U\gg\Delta$), one can apply a slight constant shift (within $5\%$)  of $t_d$ or $U$ in MGAL to outline the NRG boundaries almost perfectly in the whole $\epsilon_L/U_L-\epsilon_R/U_R$ plane. In addition, this also leads to very good predictions of the Josephson current. Nevertheless, we also discuss the limitations of the MGAL model. They can be clearly shown in the strong interaction limit. Here a much larger shift of the MGAL parameters is needed (typically $30\%$ in case of $U=10\Delta$) to faithfully capture the NRG phase boundaries. Even then, the Josephson current can differ by a factor of two from the NRG in the vicinity of the QPT. However, it is worth noting that this difference is large when compared to the precision of GAL in the half-filling or to the MGAL results for small and intermediate $U$, but still favorable when compared to other effective theories~\cite{Saldana2018}.

Because of its simplicity and reliability GAL or its modification can be used for fast and broad parameter scans like the one presented Fig.~\ref{Fig:phd_Map}. This allowed us to notice, and later confirm via NRG, several interesting properties which might be relevant for future experiments. For example, if a sufficiently large phase difference is introduced, then the doublet phase can emerge even at half-filling and for otherwise perfectly symmetric dots. In addition, the doublet phase can survive even for inter-dot hopping $t_d > \Delta$. We have also identified interesting regimes away from half-filling. For example, at intermediate $U$ and small $t_d$ an island of the doublet phase exists in the center of  the $\varepsilon_L/U_L-\varepsilon_R/U_R$ phase diagram which transits into a star-like shape with increasing $t_d$. Further increase of $t_d$ leads to broadening of the star-shape and simultaneously a small singlet island emerges in its center.

Taken together, all of the findings presented in the paper indicate that the exactly solvable GAL model and its modified version MGAL do not oversimplify the complex behavior of quantum dots coupled to superconducting leads. Instead, these effective models deliver results in good accordance with elaborate theoretical techniques such as NRG. Therefore, they can be used not only for preliminary theoretical investigations but their precision is sufficient for direct analysis of experimental data. In addition, GAL scaling can be also utilized in studies of systems that combine superconducting and normal leads via hybrid methods, where the superconducting part is threaded by AL approximation and normal part via different method, e.g., NRG ~\cite{Trocha2014,Weymann2015,Trocha2017,Domanski2017}. The quantitative agreement of GAL or MGAL with NRG results also opens a possibility that GAL or its modification may actually be an effective model of SCIAM that follows from a rigorous application of the NRG approach as discussed originally by K. G. Wilson \emph{et al.} ~\cite{Wilson1975,Krishna1980,Krishna1980II}. This would indicate a fundamental nature of the AL scalings~\eqref{eq:GAL_trans_eta}-\eqref{eq:GAL_trans_U} in the sense of approximate RG renormalizations to the corresponding parameters. We, however, leave this as an open problem for future research.

%%%%%%%%%%%%%%%%%%%%%%%%%%%%%%%%%%%%%%%%%%%%%%%%%%%%%%%%%%%
\section*{Acknowledgments}
This work was supported by the Ministry of Education, Youth and Sports of the Czech Republic through the
e-INFRA CZ (ID:90140), by the COST action CA21144 SUPERQUMAP and by Grant No. 22-22419S (M.Ž.) of the
Czech Science Foundation.
%%%%%%%%%%%%%%%%%%%%%%%%%%%%%%%%%%%%%%%%%%%%%%%%%%%%%%%%%%%
\appendix
%%%%%%%%%%%%%%%%%%%%%%%%%%%%%%%%%%%%%%%%%%%%%%%%%%%%%%%%%%%
\section{The non-interacting Green's function for SDQD}
\label{Sec:AppGF}

The AL theory for the SDQD can be derived using the non-interacting 
($U=0$) Green's function. First, we define a Nambu spinor 
$\Psi=\left(d_{L\uparrow}^{\phantom{\dag}},d_{L\downarrow}^{\dag},
d_{R\uparrow}^{\phantom{\dag}},d_{R\downarrow}^{\dag}\right)$
for the SDQD. The non-interacting, imaginary-time Nambu-Green function 
$\hat{G}_0(\tau)=-\langle \mathcal{T}_\tau[\Psi(\tau)\Psi^\dag(0)]\rangle$ 
is then a $4\times 4$ matrix,
\begin{equation}
\begin{aligned}
&\hat{G}_0(\tau)= \\
&-\begin{pmatrix}
\langle d_{L\uparrow}^{\phantom{\dag}}d_{L\uparrow}^\dag\rangle_\tau & 
\langle d_{L\uparrow}^{\phantom{\dag}}d_{L\downarrow}^{\phantom{\dag}}\rangle_\tau &
\langle d_{L\uparrow}^{\phantom{\dag}}d_{R\uparrow}^\dag\rangle_\tau & 
\langle d_{L\uparrow}^{\phantom{\dag}}d_{R\downarrow}^{\phantom{\dag}}\rangle_\tau \\[0.1em]
\langle d_{L\downarrow}^\dag d_{L\uparrow}^\dag \rangle_\tau & 
\langle d_{L\downarrow}^\dag d_{L\downarrow}^{\phantom{\dag}}\rangle_\tau & 
\langle d_{L\downarrow}^\dag d_{R\uparrow}^\dag\rangle_\tau & 
\langle d_{L\downarrow}^\dag d_{R\downarrow}^{\phantom{\dag}}\rangle_\tau \\[0.1em]
\langle d_{R\uparrow}^{\phantom{\dag}}d_{L\uparrow}^\dag\rangle_\tau & 
\langle d_{R\uparrow}^{\phantom{\dag}}d_{L\downarrow}^{\phantom{\dag}}\rangle_\tau & 
\langle d_{R\uparrow}^{\phantom{\dag}}d_{R\uparrow}^\dag\rangle_\tau & 
\langle d_{R\uparrow}^{\phantom{\dag}}d_{R\downarrow}^{\phantom{\dag}}\rangle_\tau \\[0.1em]
\langle d_{R\downarrow}^{\dag}d_{L\uparrow}^\dag\rangle_\tau & 
\langle d_{R\downarrow}^{\dag}d_{L\downarrow}^{\phantom{\dag}}\rangle_\tau & 
\langle d_{R\downarrow}^{\dag}d_{R\uparrow}^\dag\rangle_\tau & 
\langle d_{R\downarrow}^{\dag}d_{R\downarrow}^{\phantom{\dag}}\rangle_\tau \\
\end{pmatrix},
\end{aligned}
\end{equation}
where $\langle xy\rangle_\tau \equiv
\langle \mathcal{T}_\tau[x(\tau)y(0)]\rangle$.
The Green function in the Matsubara (imaginary) frequency domain reads
\begin{equation}
\begin{aligned}
\hat{G}_0(i\omega_n)&=
\int_0^\beta d\tau e^{-i\omega_n\tau}\hat{G}_0(\tau) \\
&=\left[i\omega_n\hat{I}-\hat{\varepsilon}-\hat{\Gamma}(i\omega_n)\right]^{-1},
\label{GF}
\end{aligned}
\end{equation}
where $\omega_n=(2n+1)\pi k_BT$, $\hat{I}$ is a $4\times 4$ unit matrix,
$\hat{\varepsilon}$ describes the local energy levels and hoppings 
in the isolated SDQD:
\begin{equation}
\hat{\varepsilon}=
\begin{pmatrix}
\varepsilon_{L} & 0 & -t_d & 0 \\[0.1em]
0 & -\varepsilon_{L} & 0 & t_d \\[0.1em]
-t_d & 0 & \varepsilon_{R} & 0 \\[0.1em]
0 & t_d & 0 & -\varepsilon_{R} \\
\end{pmatrix},
\end{equation}
and $\hat{\Gamma}_i(i\omega_n)$
is the hybridization function describing the coupling between the 
quantum dot $i=L,R$ and the superconducting lead:
\begin{equation}
\hat{\Gamma}(i\omega_n)=
\begin{pmatrix}
~\hat{\Gamma}_L(i\omega_n) & \hat{0} \\[0.1em]
\hat{0} & \hat{\Gamma}_R(i\omega_n)~
\end{pmatrix}
\end{equation}
with
\begin{eqnarray}
\hat{\Gamma}_j(i\omega_n)=&&
\frac{\Gamma_j}{\sqrt{\Delta^2+\omega_n^2}}
\frac{2}{\pi}\arctan\left(\frac{D}{\sqrt{\Delta^2+\omega_n^2}}\right)\nonumber\\
&&\times\begin{pmatrix}
i\omega_n & \Delta e^{i\varphi_j}\\[0.1em]
\Delta e^{-i\varphi_j} & i\omega_n
\end{pmatrix},\quad j=L,R.
\end{eqnarray}
Here $2/\pi\arctan(D/\sqrt{\Delta^2+\omega_n^2})$ is the correction due to the finite bandwidth $D$.

The Green function which corresponds to the non-interacting part of the AL Hamiltonian~\eqref{eq:DotsAL} can be then obtained by taking first the limit $D\rightarrow\infty$ and then sending $\Delta\rightarrow\infty$.
 
%%%%%%%%%%%%%%%%%%%%%%%%%%%%%%%%%%%%%%%%%%%%%%%%%%%%%%%%%%%
\section{Subspaces of the GAL Hamiltonian}
\label{Sec:AppGALH} 
While the Hilbert space of the GAL Hamiltonian is already small (only $16$ states for SDQD) it can be further 
cast into the singlet, doublet and triplet subspaces.
Following the supplementary information to Ref.~\cite{Zitko2015dd} the singlet subspace is spanned by five states:
\begin{eqnarray}
	&&
	\ket{0},\,\,
	d^\dagger_{R\downarrow}d^\dagger_{R\uparrow}\ket{0}, \,\,
		\dfrac{1}{\sqrt{2}}(d^\dagger_{L\downarrow}d^\dagger_{R\uparrow}-d^\dagger_{L\uparrow}d^\dagger_{R\downarrow})\ket{0},\nonumber \\
	&&d^\dagger_{L\downarrow}d^\dagger_{L\uparrow}\ket{0}, \,\,
	 d^\dagger_{L\downarrow}d^\dagger_{L\uparrow}d^\dagger_{R\downarrow}d^\dagger_{R\uparrow}\ket{0}
\end{eqnarray}
and its Hamiltonian reads
\begin{widetext}
	\begin{equation}
		\mathcal{H}^\text{S}=
		\begin{pmatrix}
			\frac{\tilde{U}_L+\tilde{U}_R}{2}-\tilde{\varepsilon}_L-\tilde{\varepsilon}_R &
			-\tilde{\Gamma}_Re^{-i\varphi/2} &
			0 &
			-\tilde{\Gamma}_Le^{i\varphi/2} &
			0\\
			-\tilde{\Gamma}_Re^{i\varphi/2} &
			\frac{\tilde{U}_L+\tilde{U}_R}{2}-\tilde{\varepsilon}_L+\tilde{\varepsilon}_R &
			-\sqrt{2}\tilde{t}_d &
			0&
			-\tilde{\Gamma}_Le^{i\varphi/2}\\
			0 &
			-\sqrt{2} \tilde{t}_d &
			0 &
			-\sqrt{2} \tilde{t}_d &
			0 \\
			-\tilde{\Gamma}_Le^{-i\varphi/2} &
			0 &
			-\sqrt{2} \tilde{t}_d &
			\frac{\tilde{U}_L+\tilde{U}_R}{2}+\tilde{\varepsilon}_L-\tilde{\varepsilon}_R &
			-\tilde{\Gamma}_Re^{-i\varphi/2} \\
			0 &
			-\tilde{\Gamma}_Le^{-i\varphi/2} &
			0 &
			-\tilde{\Gamma}_Re^{i\varphi/2} &
			\frac{\tilde{U}_L+\tilde{U}_R}{2}+\tilde{\varepsilon}_L+\tilde{\varepsilon}_R 
		\end{pmatrix}.
		\label{eq:Msin}
	\end{equation}
\end{widetext}

Analogously, the four doublet states
can be ordered into the doublet ket vector upon which the doublet projection of the GAL Hamiltonian yields
\begin{equation}
	d^\dagger_{L\uparrow}\ket{0},\,\,d^\dagger_{R\uparrow}\ket{0},\,\,
	d^\dagger_{L\uparrow}d^\dagger_{R\downarrow}d^\dagger_{R\uparrow}\ket{0},\,\
	d^\dagger_{L\downarrow}d^\dagger_{L\uparrow}d^\dagger_{R\uparrow}\ket{0}
\end{equation} 
which leads to:
\begin{equation}
\begin{aligned}
	&\mathcal{H}^\text{D}= \\
	&\begin{pmatrix}
		\frac{\tilde{U}_L}{2}\!-\!\tilde{\varepsilon}_L &
		-\tilde{t}_d &
		0&
		-\tilde{\Gamma}_Le^{i\varphi/2} \\
		-\tilde{t}_d &
		\frac{\tilde{U}_R}{2}\!-\!\tilde{\varepsilon}_R &
		-\tilde{\Gamma}_Re^{-i\varphi/2} &
		0 \\
		0 &
		-\tilde{\Gamma}_Re^{i\varphi/2} &
		\frac{\tilde{U}_R}{2}\!-\!\tilde{\varepsilon}_R &
		\tilde{t}_d \\
		-\tilde{\Gamma}_Le^{-i\varphi/2} &
		0 &
		\tilde{t}_d &
		\frac{\tilde{U}_L}{2}\!-\!\tilde{\varepsilon}_L
	\end{pmatrix}.
	\label{eq:Mdoub}
\end{aligned}
\end{equation}
We omit here the explicit form of the triplet state, because it never becomes the ground-state of SDQD and always yields zero eigenenergy.

In general, SDQD has to be solved numerically, which is a trivial task given the small size of the subspaces involved. Moreover, some useful limiting cases are solvable analytically.
For $\varphi=\pi$ and $\Gamma_L=\Gamma_R$ at half-filling ($\varepsilon_L=\varepsilon_R=0$) the singlet eigenvalues read
\begin{equation}
\label{eq:EVs}
\begin{aligned}
&\frac{\tilde{U}_L+\tilde{U}_R}{2},\quad\frac{\tilde{U}_L+\tilde{U}_R}{2}\pm 2\tilde{\Gamma}, \\
&\frac{\tilde{U}_L+\tilde{U}_R}{4}\pm\sqrt{\frac{(\tilde{U}_L+\tilde{U}_R)^2}{4} + 4 \tilde{t}_d^2},
\end{aligned}
\end{equation}
while doublet eigenvalues become
\begin{equation}
\label{eq:EVd}
\begin{aligned}
&\frac{1}{4}\left(\tilde{U}_L+\tilde{U}_R - 4\tilde{\Gamma}
\pm \sqrt{(\tilde{U}_L-\tilde{U}_R)^2+16\tilde{t}^2_d}\right), \\
&\frac{1}{4}\left(\tilde{U}_L+\tilde{U}_R + 4\tilde{\Gamma}
\pm \sqrt{(\tilde{U}_L-\tilde{U}_R)^2+16\tilde{t}^2_d}\right).
\end{aligned}
\end{equation}
By comparing Eqs.~\eqref{eq:EVs} and Eqs.~\eqref{eq:EVd} one can get critical values of $t_d$ that bound the $\pi$-phase region as discussed in the main text and in Sec.~\ref{Sec:AppSS}.

%%%%%%%%%%%%%%%%%%%%%%%%%%%%%%%%%%%%%%%%%%%%%%%%%%%%%%%%%%%
\section{Numerical renormalization group }
 \label{Sec:AppNRG}
  
 The NRG results presented in the paper had been calculated using 
 the open source package NRG Ljubljana~\cite{ZitkoPruschke2009,Ljubljana-code}. 
 For single channel problems, e.g., single dot at $\varphi=0$ and double-dot case with $t_d\rightarrow 0$, we used the logarithmic discretization parameter $\Lambda=2$, the maximum (minimum) number of states kept after each of the truncations was $n_s=10000$, $n_m=1000$ times the corresponding multiplicities and the cut-off energy was set to $E_C=10$ in the units of the characteristic NRG energy scale (see the manual to NRG Ljubljana \cite{Ljubljana-code}).

 For two channel problems we always used $\Lambda=4$. When calculating the profiles of sub-gap energies or the suppercurrent the remaining parameters were set as $n_s=6000$, $E_C=6$ and $n_m=1000$. Since the phase boundary calculations are less sensitive to the truncations we used $n_s=6000$, $E_C=6$ and $n_m=600$ or $n_m=1000$.
 In all cases we have used the half-bandwidth of $D=100\Delta$ which effectively suppresses band-edge related effects. For details on the derivation and implementation of the Josephson current into NRG Ljubljana see the supplementary material to Ref.~\cite{Saldana2018} and code manual~\cite{Ljubljana-code}.
 
%%%%%%%%%%%%%%%%%%%%%%%%%%%%%%%%%%%%%%%%%%%%%%%%%%%%%%%%%%%
\section{Subgap states}
\label{Sec:AppSS}

\begin{figure}[t]
	\includegraphics[width=1\columnwidth]{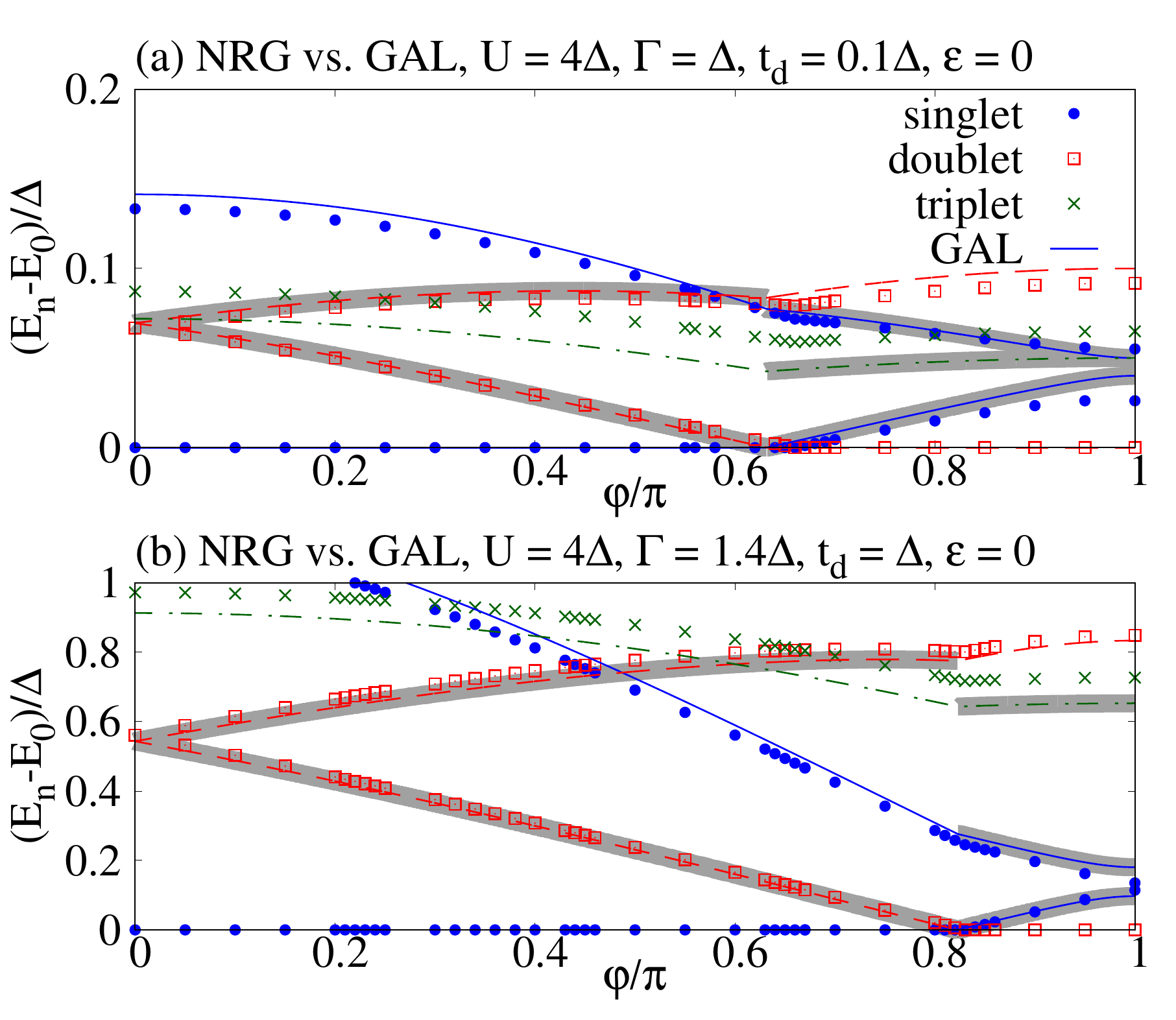}
	\caption{
		Direct comparison of NRG (symbols) and GAL (lines) results for subgap many-body states as functions of $\varphi$ calculated for two sets of parameters. Singlet states are marked by blue solid lines (GAL) and circles (NRG), doublets by red dashed lines and squares and triplets by green dot-dashed lines and crosses. Note that the differences between the energies of the excited states and the ground state energy equal the absolute values of ABS energies if the single-particle transition between the states is allowed. The ABS energies are underscored by gray stripes.}
	\label{Fig:absphi2d}
\end{figure}

In Sec.~\ref{sec:ABS}, one particularly representative case of phase-bias controlled ABS states is discussed. Two more cases are shown here in Fig.~\ref{Fig:absphi2d} for slightly larger Coulomb interaction of $U=4\Delta$ while phase-bias was kept to $\varphi=\pi$. Unlike in the main text, GAL (lines) and NRG results (points) are directly compared.

In Fig.~\ref{Fig:absphi2d}$(a)$ the case of $\Gamma=\Delta$ and $t_d=0.1\Delta$ shows a very good quantitative agreement between the GAL theory and the corresponding NRG calculations. Due to the resulting small ratio of $t_d/U$, the ABS states are pushed very close to the Fermi energy with a singlet-doublet QPT is observed at $\varphi \approx 0.6\pi$.

Setting then instead $\Gamma=1.4\Delta$ and $t_d=\Delta$ the ABS states moves the QPT to $\varphi \approx 0.8\pi$ with phase-bias controlled ABS states populating the entire gap region. The outer singlet ABS state even clearly crosses into the continuum. Once again, an overall very good quantitative agreement between the GAL theory and the corresponding NRG calculations is observed.
 
%%%%%%%%%%%%%%%%%%%%%%%%%%%%%%%%%%%%%%%%%%%%%%%%%%%%%%%%%%%
\section{Observability of doublet phase in SDQD at half-filling}
\label{Sec:AppDP}

\begin{figure}[tbh]
	\includegraphics[width=0.8\columnwidth]{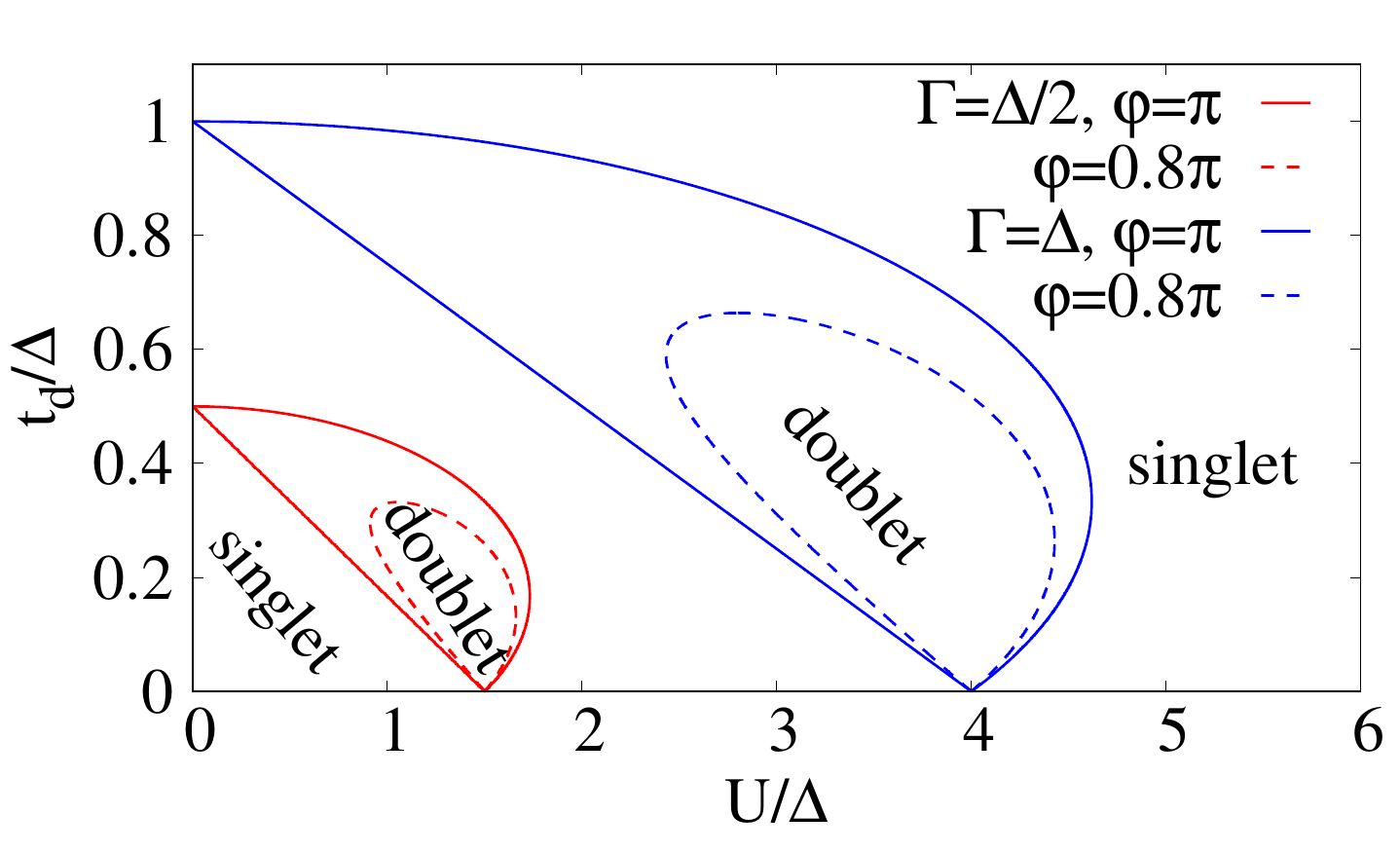}
	\caption{Phase diagrams of symmetric ($U_L=U_R$, $\Gamma_L=\Gamma_R$) half-filled case in $t_d-U$
		plane for $\varphi=\pi$ (solid lines) and $\varphi=0.8\pi$ calculated by GAL.}
	\label{Fig:phd_tdU}
\end{figure}

As discussed in the main text, formulas~\eqref{eq:GALphi1} put restrictions on the combination of parameters $U$, $\Gamma$ and $t_d$ for which the doublet phase can manifest itself at half-filling. As discussed therein and shown in Fig.~\ref{Fig:phd_tdU} in the $t_d$-$U$ plane,
the doublet phase space is largest at $\varphi=\pi$. It forms droplet-like islands surrounded by singlet phase spaces. Fig.~\ref{Fig:phd_tdU}, then clearly shows rapid shrinking of the doublet phase space with $\varphi$ as illustrated by the dashed lines for $\varphi=0.8\pi$.  

Consequently, smoothly increasing $U$ or $t_d$ at half-filling while keeping sufficiently large $\varphi$ can lead, for a proper combination of $U$ and $t_d$, to a re-entrant behavior. The system first leaves the first singlet phase and goes over into the doublet phase and then enters the second singlet phase.

%
% BibTeX users please use
%\bibliographystyle{apsrev4-1}
%\bibliography{sdqd}
%

\end{thebibliography}

\end{document}